\documentclass[11pt,oneside]{article}
\usepackage[usenames,dvipsnames]{color}
\usepackage{a4wide}
\usepackage{mathrsfs}
\usepackage{latexsym,bm}
\usepackage{graphicx}
\usepackage{indentfirst}
\usepackage{slashed}
\usepackage{amsmath}
\usepackage{amssymb}
\usepackage{color}
\usepackage{hyperref}
\usepackage{epsfig}
\usepackage[titletoc]{appendix}
\usepackage{multirow}%
\usepackage{rotating}
\usepackage{epstopdf}
\usepackage{cite}
\usepackage{placeins}
\usepackage[square,numbers,sort&compress]{natbib} 
\usepackage{tikz}
\usepackage{ulem} 
\usepackage[top=2.9cm,bottom=2.5cm,left=2.8cm,right=3cm]{geometry}
\usepackage{tabularx}

\newcommand{\email}[1]{\footnote{{\em } \texttt{#1}}}
\newcommand{\bra}{\left\langle }
\newcommand{\ket}{\right\rangle }
\newcommand{\gem}{G_{\text{EM}}}
\newcommand{\taupp}{\tau^-\to\pi^-\pi^0\nu_\tau}
\newcommand{\tauppg}{\tau^-\to\pi^-\pi^0\gamma\nu_\tau}
\newcommand{\rxt}{R\chi T}
\newcommand{\bma}{\left(\begin{matrix}}
\newcommand{\ema}{\end{matrix}\right)}

\begin{document}
\title{
\Large \bf Revisit of the electromagnetic correction to $\tau\to\pi\pi\nu_\tau$ and its implication for muon $g-2$ based on $\tau$ data }
\author{\small Zhi-Xin Li,\,\, Ao Li,\,\, Jin Hao,\, Chun-Gui Duan\email{duancg@hebtu.edu.cn},\,\,  Zhi-Hui Guo\email{zhguo@hebtu.edu.cn} \\[0.5em]
{ \small\it  Department of Physics and Hebei Key Laboratory of Photophysics Research and Application, } \\ 
{\small\it Hebei Normal University,  Shijiazhuang 050024, China}
}
\date{}

\maketitle
\begin{abstract}
In this work we focus on the evaluation of the leading-order hadronic vacuum polarization contribution from the $\pi\pi$ channel to the muon anomalous magnetic moment $a_\mu$ by using the experimental $\tau\to\pi\pi\nu_\tau$ data. The isospin breaking corrections play the decisive role in this approach of computing $a_\mu$. One of such important isospin breaking sources is the long-distance electromagnetic correction factor $G_{\rm EM}$ of the $\tau\to\pi\pi\nu_\tau$ process from the real photon radiation. The latter effect can be calculated from the $\tau\to\pi\pi\nu_\tau\gamma$ amplitude, which is revised in this work within the resonance chiral theory by simultaneously including the even-intrinsic-parity and odd-intrinsic-parity resonance operators. We update the determination of the only unknown resonance coupling through the $\omega\to\pi^0\pi^0\gamma$ decay by including contributions from both  the vector and scalar resonances. By taking other remaining contributions from the muon $g-2$ White Paper 2025, we further revise the complete value of $a_\mu$, which turns out to deviate from the newest world average result after Fermilab's measurement at the level of 2.7~$\sigma$. 
\end{abstract}

\section{Introduction}

The remarkably precise measurement of the muon anomalous magnetic moment, i.e., $a_\mu=(g-2)_\mu/2$, by Muon $g-2$ collaboration from Fermi National Accelerator Laboratory  (FNAL)~\cite{Muong-2:2021ojo,Muong-2:2023cdq,Muong-2:2025xyk}, has attracted enormous attention in particle physics community. The newest world average experimental value is $a_\mu^{\rm Exp}=116592071.5(14.5)\times 10^{-11}$~\cite{Muong-2:2025xyk}, which is in accord with the updated theoretical prediction from Standard Model (SM) $a_\mu^{\rm SM}=116592033(62)\times 10^{-11}$ released in the muon $g-2$ White Paper 2025 (WP25)~\cite{Aliberti:2025beg}. However, such an agreement is reached with a caveat: only the lattice quantum chromodynamics (QCD) numerical simulations are taken to calculate the leading order (LO) hadronic vacuum polarization (HVP) contributions to $a_\mu$. This constitutes a drastic change of strategy, comparing with that of WP20~\cite{Aoyama:2020ynm}, which relied on the data-driven dispersion relation method by taking the experimental cross sections of $e^+ e^-\to {\rm hadrons}$ as inputs. The reason behind such shift of strategy adopted in WP25 is that the new CMD-3 measurement of the $e^+e^-\to \pi^+\pi^-$ cross section~\cite{CMD-3:2023rfe,CMD-3:2023alj} is in clear tension with previous ones used in WP20, preventing reaching any definite conclusion from the existing experimental data.

In fact, among the various LO HVP contributions to $a_\mu$, the largest part comes from the $\pi\pi$ channel, which accounts for around 3/4 of the total. To resolve the discrepancies of the different $e^+e^-\to \pi^+\pi^-$ cross sections obtained from different experiments, is by no means a trivial task and will require tremendous scrutinized analyses and new measurements. Alternatively, the $\tau^-\to\pi^-\pi^0\nu_\tau$ process can provide another independent source to address the $\pi\pi$ contribution to $a_\mu$, which was first proposed in Ref.~\cite{Alemany:1997tn}. This requires us to perform the transformation from the $\pi^-\pi^0$ states in the $\tau$ decay to the $\pi^+\pi^-$ one arising in the $e^+e^-$ annihilation. To reach the subpercent precision in this transformation procedure, the isospin breaking (IB) effects must to be properly taken into account. In Ref.~\cite{Cirigliano:2002pv}, it is proposed to factorize the overall IB contributions into the product of several different pieces, including the $\pi\pi$ form factors ($F_{\pi\pi}^{(-,0)}$), kinematical factors ($\beta_{+,0}$), final-state radiation (FSR), electroweak short-distance correction ($S_{\text{EW}}$) and long-distance electromagnetic (EM) correction ($\gem$). We concentrate on the updated evaluation of the latter quantity $\gem$ in this work.

The EM correction $\gem(t)$, with $t$ the $\pi\pi$ energy squared in the $\taupp$ process, receives both contributions from the virtual photon loops and the real photon radiative effect. The pioneer calculation of the EM corrected function $G_{\text{EM}}(t)$ for the $\taupp$ process was done in Refs.~\cite{Cirigliano:2001er,Cirigliano:2002pv}, where the $\tauppg$ process was calculated by taking the minimal resonance chiral theory ($\rxt$)~\cite{Ecker:1988te} to obtain the real photon radiative correction. Later on, the vector meson dominance (VMD) model was employed in Refs.~\cite{Flores-Tlalpa:2005msx,Flores-Baez:2006yiq} to encompass more types of resonance interactions, such as the odd-intrinsic parity $\rho-\omega-\pi$ vertices, which are the main ingredients responsible for the discrepancy between the two results. The calculation of $\gem(t)$ is pushed forward in Ref.~\cite{Miranda:2020wdg} by incorporating the full sets of $\rxt$  operators that are relevant to the $O(p^6)$ low energy constants (LECs). Although the latter work includes more complete structure-dependent contributions to $\gem(t)$, the large number of unknown couplings brings in rather huge theoretical uncertainties, which turns to be one of the underlying obstacles to precisely evaluate $a_\mu$ by using the $\tau$ data~\cite{Castro:2024prg}. 

We take a different plan to pursue the calculation of $\gem(t)$ here. Instead of aiming at considering a more complete operator basis~\cite{Kampf:2011ty} as done in Ref.~\cite{Miranda:2020wdg}, we rely on the calculation of the $\tauppg$ process in Ref.~\cite{Chen:2022nxm} that is based on the odd-intrinsic parity operators of the $VVP$ and $VJP$ types in Ref.~\cite{Ruiz-Femenia:2003jdx}, apart from the minimal $\rxt$ Lagrangians~\cite{Ecker:1988te}. The merit of the $\tauppg$ calculation done in Ref.~\cite{Chen:2022nxm} is that after implementing the high-energy constraints all the unknown parameters can be fixed, except the $d_4$ coupling that is determined by the $\omega\to\pi^0\pi^0\gamma$ decay width under the on-shell approximation for the $J\omega\pi$ vertex. Therefore parameter-free predictions to the various invariant-mass distributions for the $\pi^0\pi^-,\pi^-\gamma,\pi^0\gamma$ systems and a new type of T-odd distribution are given in Ref.~\cite{Chen:2022nxm}. As a novelty, we will revise the determination of the $d_4$ coupling in this work by additionally including the scalar resonance in the $\omega\to\pi^0\pi^0\gamma$ decay. The updated value of $d_4$ will be further used in the $\tauppg$ process to evaluate the EM corrected function $\gem(t)$.  
With the new $\gem(t)$ and other IB correction terms, we update the determination of $a_\mu$ by taking the precise experimental data of the $\taupp$ process measured by the Belle~\cite{Belle:2008xpe}, ALEPH~\cite{Davier:2013sfa}, CLEO~\cite{CLEO:1999dln} and OPAL~\cite{OPAL:1998rrm} collaborations.

This paper is structured as follows. The calculation of the $\tauppg$ amplitude is recapitulated in Sec.~\ref{sec.tauppgamp}, where we also update determination of the $d_4$ resonance coupling in the $\omega\to\pi^0\pi^0\gamma$ process. The EM corrected function $\gem(t)$ by taking the revised $\tauppg$ amplitude as inputs is then computed in Sec.~\ref{sec.gem}. Next we compile the various sources of the IB corrections to the $\tau\to\pi\pi\nu_\tau$ process, and further update the evaluation of the muon anomalous magnetic moment $a_\mu$ based on the $\tau\to\pi\pi\nu_\tau$ experimental data in Sec.~\ref{sec.amu}. A short summary and conclusions are provided in Sec.~\ref{sec.concl}.  

\section{Amplitude of the $\tauppg$ process}\label{sec.tauppgamp}

In order to evaluate the EM corrected function $\gem(t)$, one will need the $\tauppg$ amplitude as inputs. Although the study of the latter process based on the resonance operator basis of Refs.~\cite{Ecker:1988te,Ruiz-Femenia:2003jdx} has been presented in detail in Ref.~\cite{Chen:2022nxm}, we recapitulate the main formulas here for self-containedness and also for establishing the notations. 
The amplitude of the $\tau^{-}(P) \to \pi^{-}(p_{1})+\pi^{0}(p_{2})+\nu_{\tau}(q)+\gamma(k)$ decay can generally be written as~\cite{Cirigliano:2002pv} 
\begin{align}\label{eq.Amplitude-t}
\mathcal{M}=eG_{F}V_{ud}^{*}\varepsilon^{u}(k)^{*}
&\left\lbrace 
F_{\nu}\bar{u}(q)\gamma^{\nu}(1-\gamma_{5})(m_{\tau}+\slashed{P}-\slashed{k})\gamma_{\mu}u(P)+ \right.\\\notag
&+\left.
(V_{\mu\nu}-A_{\mu\nu})\bar{u}(q)\gamma^{\nu}(1-\gamma_{5})u(P)
\right\rbrace \,,
\end{align}
where the $F_{\nu}$ term includes the effect of the bremsstrahlung off the $\tau$ lepton  
\begin{align}
F_{\nu}=\left(p_{2}-p_{1}\right)_{\nu}F_{\pi\pi}^{(-)}(t)/\left(2P\cdot k\right) \,,
\end{align} 
with $F_{\pi\pi}^{(-)}(t)$ the charged pion vector form factor governing the nonradiative two-pion tau decay and $t=(p_1+p_2)^2$. 
The hadronic tensor amplitude $V_{\mu\nu}$ takes the form 
\begin{align}\label{eq.Teson-V}
V_{\mu\nu}=&F_{\pi\pi}^{(-)}(u)\frac{p_{1\mu
}}{p_{1}\cdot k }\left(p_{1}+k-p_{2}\right)_{\nu}-F_{\pi\pi}^{(-)}(u)g_{\mu\nu} + \frac{F_{\pi\pi}^{(-)}(u)-F_{\pi\pi}^{(-)}(t)}{\left(p_{1}+p_{2}\right)\cdot k}\left(p_{1}+p_{2}\right)_{\mu}\left(p_{2}-p_{1}\right)_{\nu}+ \notag \\
&+
v_{1}\left(g_{\mu\nu}p_{1}\cdot k - p_{1\mu}k_{\nu}\right)+
v_{2}\left(g_{\mu\nu}p_{2}\cdot k - p_{2\mu}k_{\nu}\right)+
v_{3}\left(p_{1\mu}p_{2}\cdot k - p_{2\mu}p_{1}\cdot k\right)p_{1\nu}+\notag \\
&+
v_{4}\left(p_{1\mu}p_{2}\cdot k - p_{2\mu}p_{1}\cdot k\right)\left(p_{1}+p_{2}+k\right)_{\nu}\,,
\end{align}
which comprises both structure-independent (SI) part, such as the terms with $F_{\pi\pi}^{(-)}(t)$ and $F_{\pi\pi}^{(-)}(u)$, being $u=(P-q)^{2}$, and the structure-dependent(SD) part, such as the $v_{i}$ terms. The $A_{\mu\nu}$ amplitude reads 
\begin{align}\label{eq.Teson-A}
A_{\mu\nu}=i\left(
a_{1}\epsilon_{\mu\nu\rho\sigma}p_{1}^{\rho}k^{\sigma}+a_{2}\epsilon_{\mu\nu\rho\sigma}p_{2}^{\rho}k^{\sigma}
+a_{3}p_{1\nu}\epsilon_{\mu\rho\beta\sigma}k^{\rho}p_{1}^{\beta}p_{2}^{\sigma}
+a_{4}p_{2\nu}\epsilon_{\mu\rho\beta\sigma}k^{\rho}p_{1}^{\beta}p_{2}^{\sigma}
\right)\,,
\end{align}
which only contains the SD terms.

The $V_{\mu\nu}$ and $A_{\mu\nu}$ amplitudes of Ref.~\cite{Cirigliano:2002pv} are calculated by taking the minimal $\rxt$ Lagrangian~\cite{Ecker:1988te} and the LO Wess-Zumino-Witten term~\cite{Wess:1971yu,Witten:1983tw} that are complete at $O(p^4)$ in the low-energy regime. Later on, the VMD calculation~\cite{Flores-Tlalpa:2005msx} shows that the narrow vector-meson $\omega$ via the anomalous $\rho\omega\pi$ interaction can give noticeable effect in the $\tauppg$ process. The full $\rxt$ operators relevant to the $O(p^6)$ in the bases of Refs.~\cite{Cirigliano:2006hb,Kampf:2011ty} are recently employed to improve the $\tauppg$ amplitude in Ref.~\cite{Miranda:2020wdg}. However, due to the vast unknown resonance couplings, rather large uncertainties are assigned in the latter reference. A variant calculation to amend the $\tauppg$ amplitude by using the odd-intrinsic parity $\rxt$ operator basis of Ref.~\cite{Ruiz-Femenia:2003jdx} is carried out in Ref.~\cite{Chen:2022nxm}, so that the contribution of the narrow vector-meson $\omega$ resonance is included within the chiral theory. 
It is mentioned that the odd-intrinsic parity $\rxt$ operators of Ref.~\cite{Ruiz-Femenia:2003jdx} with the extension by incorporating $\eta$ and $\eta'$~\cite{Chen:2012vw} have been demonstrated to be very successful in many phenomenological studies, such as the radiative processes of $V\to P\gamma^{(*)}$, $P\to V\gamma^{(*)}$, $e^+e^-\to K K^*$, $\tau\to  P(V,\gamma)\nu_\tau$~\cite{Chen:2013nna,Chen:2014yta,Yan:2023nqz,Guo:2008sh,Guo:2010dv}. 
A thorough phenomenological discussion of the $\tauppg$ is given in Ref.~\cite{Chen:2022nxm}, including the spectra of the photon energy $E_\gamma$, invariant-mass distributions of the $\pi^-\gamma$, $\pi^0\gamma$ and $\pi^-\pi^0$ systems, and the interesting T-odd distribution. 
In this work, we further exploit the $\tauppg$ amplitude in Ref.~\cite{Chen:2022nxm} to calculate the EM corrected function $\gem(t)$ for the $\taupp$ process that will be addressed in detail in next section. We recapitulate the calculation of the $\tauppg$ amplitude below. Furthermore, we revise the determination of the $d_4$ parameter by including the scalar resonance effect in the $\omega\to\pi^0\pi^0\gamma$ decay, in addition to the vector exchanges already considered in Ref.~\cite{Chen:2022nxm}. 

The minimal $\rxt$ interaction Lagrangian, which includes the vector and axial-vector resonances, are given by  
\begin{align}\label{eq.lagv2}
&\mathcal{L}_{V}=\frac{F_{V}}{2\sqrt{2}}\left\langle
V_{\mu\nu}f_{+}^{\mu\nu}
\right\rangle 
+
i\frac{G_{V}}{\sqrt{2}}
\left\langle 
V_{\mu\nu}u^{\mu}u^{\nu}
\right\rangle ,
\\
&\mathcal{L}_{A}=\frac{F_{A}}{2\sqrt{2}}
\left\langle 
A_{\mu\nu}f_{-}^{\mu\nu}
\right\rangle\,,\label{eq.laga2}
\end{align}
where the antisymmetric tensor formalism is used for the vector and axial-vector resonances~\cite{Ecker:1988te}. To incorporate additional contributions arising from hadronic interaction vertices of the $\rho\pi\gamma$, $\omega\pi\gamma$, and $\omega\rho\pi$ types, the $VJP$ and $VVP$ Lagrangians within the framework of $\rxt$~\cite{Ruiz-Femenia:2003jdx} are employed: 
\begin{align}\label{eq.VVP}
\mathcal{L}_{VVP}=& \quad d_{1}\epsilon_{\mu\nu\rho\sigma}\left\langle
\left\{
V^{\mu\nu},V^{\rho\alpha}
\right\}\nabla_{\alpha}u^{\sigma}
\right\rangle
+ id_{2}\epsilon_{\mu\nu\rho\sigma}
\left\langle
\left\{
V^{\mu\nu},V^{\rho\sigma}
\right\}\chi_{-}
\right\rangle \notag \\
&\quad
+
d_{3}\epsilon_{\mu\nu\rho\sigma}\left\langle
\left\{
\nabla_{\alpha}V^{\mu\nu},V^{\rho\alpha}
\right\}u^{\sigma}
\right\rangle
+d_{4}\epsilon_{\mu\nu\rho\sigma}\left\langle
\left\{
\nabla^{\sigma}V^{\mu\nu},V^{\rho\alpha}
\right\}u_{\alpha}
\right\rangle ,
\end{align}
and 
\begin{align}\label{eq.VJP}
&\mathcal{L}_{VJP}= \frac{c_{1}}{M_{V}}\epsilon_{\mu\nu\rho\sigma}
\left\langle
\left\{
V^{\mu\nu},f_{+}^{\rho\alpha}
\right\}\nabla_{\alpha}u^{\sigma}
\right\rangle 
+\frac{c_{2}}{M_{V}}\epsilon_{\mu\nu\rho\sigma}\bra \left\{
V^{\mu\alpha},f_{+}^{\rho\sigma}
\right\} \nabla_{\alpha}u^{\nu} \ket \notag \\
&+
\frac{ic_{3}}{M_{V}}\epsilon_{\mu\nu\rho\sigma}
\bra
\left\{
V^{\mu\nu},f_{+}^{\rho\sigma}
\right\}\chi_{-}
\ket
+\frac{ic_{4}}{M_{V}}\epsilon_{\mu\nu\rho\sigma}
\bra
V^{\mu\nu}
\left[
f_{-}^{\rho\sigma},\chi_{+}
\right]
\ket 
+\frac{c_{5}}{M_{V}}\epsilon_{\mu\nu\rho\sigma}
\bra
\left\{
\nabla_{\alpha}V^{\mu\nu},f_{+}^{\rho\alpha}
\right\}u^{\sigma}
\ket \notag \\
&
+\frac{c_{6}}{M_{V}}\epsilon_{\mu\nu\rho\sigma}
\bra
\left\{
\nabla_{\alpha}V^{\mu\alpha},f_{+}^{\rho\sigma}
\right\}u^{\nu}
\ket
+\frac{c_{7}}{M_{V}}\epsilon_{\mu\nu\rho\sigma}
\bra
\left\{
\nabla^{\sigma}V^{\mu\nu},f_{+}^{\rho\alpha}
\right\}u_{\alpha}
\ket ,
\end{align}
where $V$ and $P$ correspond to the vector resonance and light pseudoscalar meson, respectively, and $J$ denotes the external source field. The Feynman diagrams that contribute to the vector ($v_{i=1\cdots 4}$) and axial-vector ($a_{i=1\cdots 4}$) form factors are illustrated in Ref.~\cite{Chen:2022nxm}. Their final expressions are given as the sum of two parts
\begin{align}
v_{i}=v_{i}^{\text{CEN}}+v_{i}^{VVP},\quad 
a_{i}=a_{i}^{\text{CEN}}+a_{i}^{VVP},
\end{align}
where $v_{i}^{\text{CEN}}$ and $a_{i}^{\text{CEN}}$ are the contributions by the minimal resonance chiral Lagrangians, and $v_{i}^{VVP}$ and $a_{i}^{VVP}$ encode the contributions from the interactions in Eqs.~\eqref{eq.VVP} and \eqref{eq.VJP}. Their explicit expressions can be found in Refs.~\cite{Cirigliano:2002pv,Chen:2022nxm}. 

By imposing the QCD short-distance constraints to the form factors and Green functions, one can effectively set relations to resonance coupling constants~\cite{Chen:2012vw,Chen:2014yta,Ruiz-Femenia:2003jdx,Guo:2010dv,Ecker:1989yg,Cirigliano:2006hb,Roig:2013baa}, which also guarantees that the theory will not produce behaviors contradicting QCD in the high energy limit. The pertinent short-distance relations of the resonance coupling constants in our study read 
\begin{align}\label{eq.helecs1}
&c_{1}+4c_{3}=0, \qquad c_{1}-c_{2}+c_{5}=0, \qquad c_{5}-c_{6}=\frac{N_{C}M_{V}}{64\sqrt{2}\pi^2}F_{V}\,,  \nonumber\\ 
&d_{1}+8d_{2}=-\frac{N_{C}M_{V}^{2}}{(8\pi F_{V})^{2}} +\frac{F^2}{4F_{V}^{2}}\,,\qquad
d_{3}=-\frac{N_{C}M_{V}^{2}}{(8\pi F_{V})^{2}} +\frac{F^2}{8F_{V}^{2}}\,,
\end{align}
which is valid in the leading $1/N_C$ expansion. While, in practice parts of the subleading $1/N_C$ effects are also required to be considered, such as the finite decay widths $\Gamma_R$ in the resonance propagators, as done in many phenomenological studies based on $\rxt$~\cite{
Cirigliano:2002pv,Miranda:2020wdg,Chen:2022nxm,Chen:2013nna,Chen:2014yta,Yan:2023nqz,Guo:2008sh,Guo:2010dv}. Though the finite  widths of resonances belong to the subleading $1/N_C$ terms, the proper inclusion of the finite-width effects can be crucial, since they will greatly affect the invariant-mass distributions. In this work we follow Ref.~\cite{Chen:2022nxm} to also incorporate the finite widths for the pertinent resonances in the $\tauppg$ amplitudes.
After taking the above constraints of Eq.~\eqref{eq.helecs1} and the on-shell approximation for the $VJP$ vertices, only the parameter $d_{4}$ in Eq.~\eqref{eq.VVP} remains unknown. In Ref.~\cite{Chen:2022nxm} the value of $d_{4}$ was determined from the $\omega \to \pi^0 \pi^0 \gamma$ decay by only including the contribution from the vector resonances. The various observables in $\tauppg$ are found to be sensitive to $d_4$. As an improvement, we further include the contributions from scalar resonances to the $\omega\to\pi^0\pi^0\gamma$ decay, in addition to the vector-meson resonance exchanges considered in Ref.~\cite{Chen:2022nxm}, in order to give a more realistic determination of $d_4$.

For the $\omega(q) \to \pi^0(p_1) \pi^0(p_2) \gamma(k)$ process, the contribution from the vector resonance exchange to its amplitude has been computed in Ref.~\cite{Chen:2022nxm}, and the explicit expression is 
\begin{align}\label{eq.amw}
&    \mathcal{M} _{\omega(q) \to \pi^0(p_1) \pi^0(p_2) \gamma(k)}^{\text{vector}}=\frac{2}{F}  \bigg\{ d_{1}   ( \epsilon _{\lambda \delta \mu \sigma}p_{1\nu }p_{1}^{\sigma }+ \epsilon _{\mu \nu \lambda \sigma}p_{1\delta }p_{1}^{\sigma }    )
    +4d_{2}m_{\pi}^{2}\epsilon _{\mu \nu \lambda \delta }  \nonumber\\
    &+d_{3}  [ \epsilon _{\lambda \delta \mu \sigma }  ( k+p_{2}    )_{\nu } p_{1}^{\sigma }-\epsilon _{\mu \nu \lambda \sigma } q_{\delta } p_{1}^{\sigma }    ] 
  +d_{4}  [ \epsilon _{\lambda \delta \mu \sigma }  ( k+p_{2}    )^{\sigma } p_{1 }^{\nu}-\epsilon _{\mu \nu \lambda \sigma } q^{\sigma} p_{1\delta}   ]  \bigg\}\nonumber\\
    &\times D^{\lambda \delta ,\beta \theta }\left ( k+p_{2},m_{\rho}^{2}   \right ){g_{\rho } }\epsilon _{\beta \theta \xi \alpha } k^{\xi } \varepsilon_{\gamma}^{\ast\alpha}(k) \frac{1}{m_{\omega } }\big[ q^{\mu }\varepsilon_{\omega}^{\nu}\left ( q \right )-q^{\nu }\varepsilon_{\omega}^{\mu}\left ( q \right ) \big]
    +\bigg( p_{1}\leftrightarrow p_{2} \bigg)\,,
\end{align}
where the on-shell $\rho\pi\gamma$ coupling is 
\begin{align}\label{eq.gkstar0}
g_{\rho}&=\frac{\sqrt{2}e }{3M_{V}F}\left [ \left ( c_2-c_1+c_5-2c_6 \right )m_{\rho}^2 +\left ( c_1+c_2+8c_3-c_5 \right )m_{\pi}^2 \right]\,,
\end{align}
and the $\rho$ propagator in the antisymmetric tensor formalism reads 
\begin{align}
D^{\mu\nu,\rho\sigma}(k,M_{V}^2)=\frac{1}{M_{V}^2D_{\rho}(k^2)}\left \{ g^{\mu\rho}g^{\nu\sigma}(M_{V}^{2}-k^{2})+g^{\mu\rho}k^{\nu}k^{\sigma}-g^{\mu\sigma}k^{\nu}k^{\rho}-
(\mu\leftrightarrow \nu) \right \}\,,
\end{align}
with 
\begin{align} \label{eq.Dpro}
D_{\rho}(k^2)= M_{\rho}^{2}-k^2-iM_{\rho}\Gamma_{\rho}(k^2) \,,
\end{align}
and 
\begin{align}
\Gamma_{\rho}(k^2)=\frac{M_{\rho}k^2}{96\pi F_{\pi}^{2}}
\left[
\left(
1-\frac{4m_{\pi}^{2}}{k^2}
\right)^{\frac{3}{2}}
\theta\left(
k^2-4m_{\pi}^{2}
\right)+\frac{1}{2}
\left(
1-\frac{4m_{K}^{2}}{k^2}
\right)^{\frac{3}{2}}\theta\left(
k^2-4m_{K}^{2}
\right)
\right]\,.
\end{align}

\begin{figure}[htbp]
\centering
\includegraphics[width=1\textwidth,angle=-0]{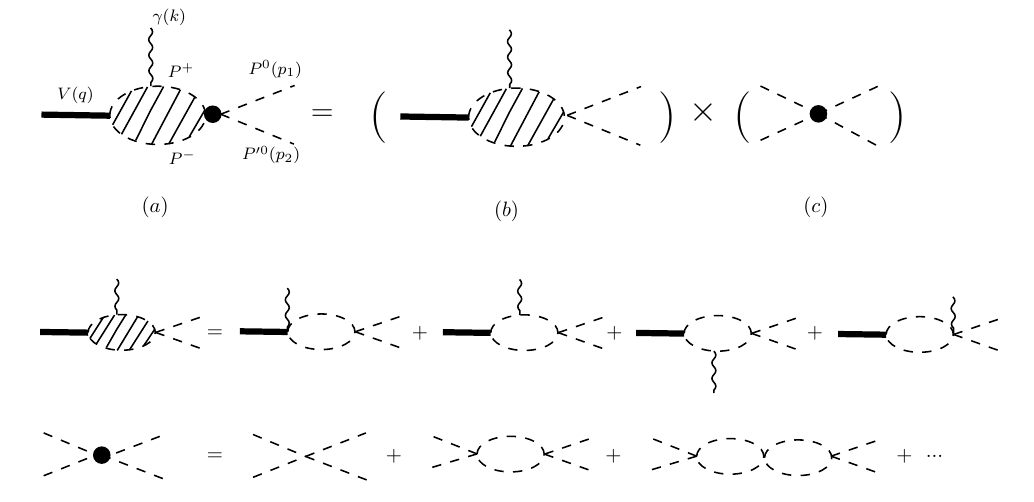}
\caption{ \quad Illustration of relevant Feynman diagrams for the scalar resonance contribution to the $\omega\to\pi^0\pi^0\gamma$ decay. The thick solid black line and the wiggle line denote the $\omega$ resonance and the photon, respectively, while the dashed lines stand for the pions.} \label{fig.loop2}
\end{figure}

In addition to the vector exchanges, different approaches have been proposed in literature, such as the linear sigma model~\cite{Bramon:2001un,Escribano:2006mb,Oh:2003zz}, phenomenological meson exchange models~\cite{Gokalp:2000xy,Eidelman:2010ta}, unitarized chiral amplitude~\cite{Oller:1998ia,Oller:2002na,Palomar:2001vg}, to include the contributions of the scalar resonances to the $\omega \to \pi^0 \pi^0 \gamma$ process. In this work we will exploit the unitarized chiral approach to account for the contributions from the scalar resonances in the $\omega \to \pi^0 \pi^0 \gamma$ decay. In this approach, the scalar resonances are incorporated through the rescattering of the charged mesons in the loops into the neutral meson pair. The chiral loop diagrams with charged mesons as the intermediate states are illustrated in Fig.~\ref{fig.loop2}. For the $\omega \to \pi^0 \pi^0 \gamma$ process, the intermediate $K^+K^-$ loop plays the predominant role, since the $\omega\pi^+\pi^-$ coupling is severely suppressed by isospin breaking. Therefore we will neglect this latter tiny effect.  

By taking the $\rxt$ Lagrangian in Eq.~\eqref{eq.lagv2}, one can calculate all the four Feynman diagrams shown in the second row of Fig.~\ref{fig.loop2} within the dimensional regularization. It turns out that the sum of these four diagrams will become finite and free of ultraviolet (UV) divergences when imposing the relation of $F_V = 2G_V$, which confirms the findings in Refs.~\cite{Oller:1998ia,Oller:2002na,Palomar:2001vg}. It is interesting to point out the latter relation $F_V = 2G_V$ coincides with the high-energy constraints used in Ref.~\cite{Cirigliano:2002pv}
\begin{align}\label{eq.LECs-1}
F_{A}=F_{\pi},\quad F_{V}=\sqrt{2}F_{\pi}, \quad G_{V}=F_{\pi}/\sqrt{2}\,, 
\end{align}
which is also one of the two sets of constraints employed in Ref.~\cite{Chen:2022nxm}. The finiteness of the chiral loop diagrams for the $\omega\to\pi^0\pi^0\gamma$ process presents an attractive feature, since one does not need to introduce the local operators to renormalize this amplitude. Therefore we will stick to the high-energy constraints in Eq.~\eqref{eq.LECs-1} in this work and refrain from discussing the other set of constraint with $F_V=3 G_V$ analyzed in Ref.~\cite{Chen:2022nxm}. By summing the contributions from the four diagrams in the second row of Fig.~\ref{fig.loop2}, implementing the relation $F_V=2G_V$ and including the scalar resonance effects via the resummation of the light pesudoscalar loops as illustrated in the third row of Fig.~\ref{fig.loop2}, the following gauge-invariant amplitude results 
\begin{align}\label{eq.ampwppgscalar}
    \mathcal{M}_{\omega(q) \to \pi^0(p_1) \pi^0(p_2) \gamma(k)}^{\text{scalar}}=-\frac{ e G_V M_\omega}{4\pi^2 F_{\pi}^2 m_{K^+}^{2}}I(a,b)[(q \cdot k)(\varepsilon_{\gamma}^{\ast} \cdot \varepsilon_{\omega})-(q \cdot\varepsilon_{\gamma}^{\ast})(k \cdot \varepsilon_{\omega})]t_{K^+ K^- \to \pi^0 \pi^0}^{S}\,, 
\end{align}
where the UV finite loop function $I(a,b)$ reads~\cite{Close:1992ay} 
\begin{align}\label{eq.Iab}
    I(a,b)=\frac{1}{2(a-b)}-\frac{2}{(a-b)^2}\left[ f\left(\frac{1}{b}\right)-f\left(\frac{1}{a}\right)\right]+\frac{a}{(a-b)^2}\left[ g\left(\frac{1}{b}\right)-g\left(\frac{1}{a}\right) 
    \right]\,,
\end{align}
with 
\begin{align}
  a=&{m_{\omega}^2}/m_{K^+}^2 \,, \qquad  b= s/m_{K^+}^2 \,,\nonumber\\
    f(z) &=
    \begin{cases}
        -\left[\arcsin\left(\frac{1}{2\sqrt{z}}\right)\right]^2 & z > \frac{1}{4}\\
        \frac{1}{4}\left(\log\frac{\eta_ +}{\eta_-}- i\pi\right)^2 & z < \frac{1}{4}
    \end{cases}\,,  \qquad  \bigg( \eta_{\pm}=\frac{1 \pm \sqrt{1-4z}}{2} \bigg) \nonumber\\
    g(z) &=
    \begin{cases}
        \sqrt{4z - 1}\arcsin\left(\frac{1}{2\sqrt{z}}\right) & z > \frac{1}{4}\\
        \frac{1}{2}\sqrt{1 - 4z}\left(\log\frac{\eta_+}{\eta_-}- i\pi\right) & z < \frac{1}{4}
    \end{cases}\,, 
\end{align}
and $t_{K^+ K^- \to \pi^0 \pi^0}^{S}$ denotes the unitarized $S$-wave $K^+ K^- \to \pi^0 \pi^0$ scattering amplitude. The scalar resonance enters in the isoscalar channel, while the magnitude of the nonresonant isotensor amplitude with $I=2$ is much smaller, which will be neglected here. 
The $S$-wave $t_{K^+ K^- \to \pi^0 \pi^0}^S$ amplitude is related to that of $(I,J)=(0,0)$ channel through 
$t_{K^+ K^- \to \pi^0 \pi^0}^S=t_{K \bar{K},\pi\pi}^{IJ=00}/\sqrt{3}$. 
For the $\omega\to\pi^0\pi^0\gamma$ process, we only need the scattering amplitude up to the $\omega$ mass, which is much below the first inelastic $K\bar{K}$ threshold. It has been demonstrated in many works~\cite{Oller:1997ti,Oller:1998hw,Oller:1998zr} that for the $(I,J)=(0,0)$ case it is enough to take the leading-order chiral meson-meson scattering amplitude in the unitarization procedure to well describe the scattering phase shifts in the energy region below $M_\omega$. The unitarized $(I,J)=(0,0)$ amplitude with $\pi\pi$ and $K\bar{K}$, labeled as channels 1 and 2 in order, takes the form~\cite{Oller:1998zr,Guo:2011pa,Guo:2012yt,Gao:2019idb}
\begin{equation}
 \mathcal{T}^{IJ=00}(s)= \big[\, 1-N(s)\cdot G(s)\, \big]^{-1}\cdot N(s)\,,
\end{equation}
with 
\begin{eqnarray}
N(s) =   \bma
      N_{11}(s)  & N_{12}(s)  \\
      N_{21}(s)  & N_{22}(s)  \\
   \ema\,, \qquad 
G(s) =   \bma
G_{1}(s)  & 0 \\
      0  & G_{2}(s)  \\
   \ema\,. 
\end{eqnarray}
The LO $S$-wave chiral amplitudes $N_{ij}$ are~\cite{Gasser:1984gg} 
\begin{align}\label{eq.n00}
N_{11}=\frac{2s-m_{\pi}^{2}}{2F_{\pi}^{2}}\,,\quad
N_{12}=N_{21}=\frac{\sqrt{3}s}{4F_{\pi}^{2}}\,, \quad 
N_{22}=\frac{3s}{4F_{\pi}^{2}}\,. 
\end{align}
The analytical expression of the function $G_{j}(s)$ is given by~\cite{Oller:1998zr}
\begin{align}\label{eq.Gs1}
    G_{j}(s)=-\frac{1}{16\pi^2}
    \left[a_{SL,j}(\mu^2)+\log\frac{m_{2,j}^2}{\mu^2}-x_+\log\frac{x_+-1}{x_+}-x_-\log\frac{x_--1}{x_-}\right],
\end{align}
with 
\begin{equation}
 x_{\pm}=\frac{s+m_{1,j}^2-m_{2,j}^2}{2s}\pm\frac{q(s)}{\sqrt{s}}\,,\qquad q(s)=\frac{\sqrt{[s-(m_{1,j}+m_{2,j})^2][s-(m_{1,j}-m_{2,j})^2]}}{2\sqrt{s}}\,,\label{eq.Gs3}
\end{equation}
and $m_{1,j}$ and $m_{2,j}$ the masses of the two particles in channel $j$. The subtraction constant $a_{SL,j}$ at $\mu = 770$ MeV in Eq.~\eqref{eq.Gs1} has been determined by fitting experimental phase shifts and we directly take their values from Ref.~\cite{Gao:2019idb}, which also uses the LO chiral amplitudes in the unitarization procedure. For the $(I,J)=(0,0)$ channel, the values of the two subtraction constants are $a_{SL,1}=-1.13$ for the $\pi\pi$ channel and $a_{SL,2}=-1.93$ for the $K\bar{K}$ channel. It has been verified~\cite{Gao:2019idb} that the resulting $\pi\pi$ phase shifts in such formalism can well reproduce experimental data and Roy equation results. 
With such inputs, the resulting pole position of the $f_0(500)$/$\sigma$ scalar resonance $\sqrt{s}_\sigma=(0.465-i0.237)$~GeV, is also compatible with the dispersive result in Ref.~\cite{Pelaez:2015qba}. These facts imply that our treatment of the scalar resonance effects in the $\omega\to\pi^0\pi^0\gamma$ decay is reliable.

The explicit expression for the unitarized $K^+K^-\to\pi^0\pi^0$ amplitude in Eq.~\eqref{eq.ampwppgscalar} is then given by 
\begin{align}\label{eq.Gs}
t_{K^+ K^- \to \pi^0 \pi^0}^{S}(s)= \frac{\mathcal{T}_{21}^{IJ=00}(s)}{\sqrt 3}=\frac{N_{12}/\sqrt 3}{1-G_{2}(s) N_{22}-G_{1}(s) [N_{11}+G_{2}(s)N_{12}^{2}-G_{2}(s) N_{11} N_{22}]}\,. 
\end{align}

By including both contributions from vector and scalar resonances, the full expression for the decay width of the $\omega(q)\to\pi^0(p_1)\pi^0(p_2)\gamma(k)$ process reads 
\begin{align}\label{eq.Gs11}
    \Gamma_{{\omega} \to {\pi^0} {\pi^0} \gamma} &=\frac{1}{2}\int_{4m_\pi^2}^{M_\omega^2} d s \int_{t_{-}}^{t_{+}}dt
    \frac{1}{(2\pi)^3}\frac{1}{32m_{\omega}^3}\frac{1}{3}\bigg|\mathcal{M} _{\omega \to \pi^0 \pi^0 \gamma}^{\text{vector}}+\mathcal{M} _{\omega \to \pi^0 \pi^0 \gamma}^{\text{scalar}}\bigg|^2 \,,
\end{align}
with  
\begin{eqnarray}
 s= (p_1+p_2)^2\,, \qquad t=(k+p_2)^2\,,
\end{eqnarray}
and   
\begin{eqnarray}
t_{\pm}&=& \frac{M_\omega^2+2m_\pi^2-s}{2}\pm \frac{\sqrt{s^2-4m_\pi^2s}(M_\omega^2-s)}{2s}\,. 
\end{eqnarray}
It is pointed out that a missing $1/2$ factor in the phase space integral of the $\omega\to\pi^0\pi^0\gamma$ decay in Ref.~\cite{Chen:2022nxm} is now corrected here. After imposing the high-energy constraints in Eqs.~\eqref{eq.helecs1} and \eqref{eq.LECs-1}, $d_4$ turns out to be the only unknown resonance coupling, and we determine its value by using the Particle Data Group (PDG)\cite{ParticleDataGroup:2024cfk} result for the $\omega\to\pi^0\pi^0\gamma$ decay width, which is   
\begin{align}\label{eq.brb}
\Gamma_{\omega \to \pi^0 \pi^0 \gamma}^{\text{Exp}}=(5.8 \pm 1.0) \times {10^{-4}}~{\rm MeV}\,. 
\end{align}
Two solutions for the parameter $d_{4}$ are obtained 
\begin{eqnarray}\label{eq.d4n}
d_{4}=& -0.42\pm 0.07\,, \\
d_{4}=& 1.01\pm 0.07\,. \label{eq.d4p}
\end{eqnarray}
In later discussions, we will designate the negative solution of $d_4$ in Eq.~\eqref{eq.d4n} as Sol-A and the positive one in Eq.~\eqref{eq.d4p} as Sol-B. 
Both solutions of $d_4$ will be taken to calculate the EM corrected function in the following section. 
Moreover, we also quantify to what the extent the inclusion of scalar resonance effects in $\omega\to\pi^0\pi^0\gamma$ can modify the determination of $d_4$. For this purpose, we recompute $d_4$ in the pure-vector scenario (as mentioned previously a mistake of Ref.~\cite{Chen:2022nxm} has been corrected here) to neglect the scalar contribution in the $\omega\to\pi^0\pi^0\gamma$ amplitude. This provides the solutions with $d_4^{(V)} =-0.38 \pm 0.07 $ and $1.06\pm 0.07$, to be compared with our preferred ones in Eqs.~\eqref{eq.d4n} and \eqref{eq.d4p} that are obtained by including both the vector and scalar resonances. The corresponding relative shifts are about $10\%$ (Sol-A) and $4\%$ (Sol-B), while the absolute shifts of the central values of the two solutions are  similar: $|\Delta d_4|= 0.04$ (Sol-A) and $0.05$ (Sol-B).

\section{Electromagnetic corrections to $\taupp$} \label{sec.gem}

The EM corrections to the $\taupp$ process consist of two parts, the virtual correction from the photon loops and the real photon radiative correction. The virtual correction from the one-loop photon diagrams to the $\taupp$ decay has been computed in Ref.~\cite{Cirigliano:2001er}. Up to $O(\alpha)$, being $\alpha=e^2/(4\pi)$ the fine structure constant of quantum electrodynamics (QED), the real photon radiative correction can be calculated from the $\tauppg$ process. 
A novelty of the present study stems from the use of the revised $\tauppg$ amplitude discussed in the previous section  to calculate the real radiative correction part of $\gem(t)$. 
In this way, our work also provides an independent crosscheck of the previous calculations~\cite{Cirigliano:2002pv,Miranda:2020wdg}.

In the case of neglecting the EM corrections, the differential decay width of $\taupp$ can be written as  
\begin{align}\label{eq.nonraddw}
\frac{d\Gamma^{(0)}_{\tau_{\pi\pi}}}{dt}  =
\frac{G_{F}^{2}\left|V_{ud}\right|^{2}m_{\tau}^{3}S_{\text{EW}}}{384\pi^{3}}
\left( 1-\frac{4m_{\pi}^{2}}{t} \right)^{\frac{3}{2}} \left( 1-\frac{t}{m_{\tau}^2} \right)^{2}\left(1+\frac{2t}{m_{\tau}^{2}}
\right)
\left|F_{\pi\pi}^{(-)}(t)\right|^{2} \,,
\end{align}
which is governed by the $\pi^-\pi^0$ vector form factor $F_{\pi\pi}^{(-)}(t)$, depending on a single variable $t=(p_{\pi^-}+p_{\pi^0})^2$. The superscript (0) in $d\Gamma^{(0)}_{\tau_{\pi\pi}}/dt$ denotes the quantify that does not include the EM corrections. The short-distance electroweak correction $S_{\text{EW}}=1.0233\pm 0.0003$ from Ref.~\cite{Castro:2024prg} will be used throughout. 

When including the effects from the virtual photons, one would need to introduce another kinematical variable $u=(p_{\tau^-}-p_{\pi^-})^2$, in addition to $t$, to calculate the EM contributions.  
Up to $O(\alpha)$, the double differential decay width of the photon inclusive process takes the form~\cite{Cirigliano:2001er}
\begin{align}\label{eq.dpipiincl}
\frac{d\Gamma_{\tau_{\pi\pi[\gamma]}}}{dt\,du}=\frac{G_{F}^{2}  S_{\text{EW}} |V_{ud}|^{2}}{64 \pi^{3} m_{\tau}^{3}}
\left|F_{\pi\pi}^{(-)}(t) \right|^{2} D(t,u)
\left[ 
1+2f^{\text{elm}}_{\text{loop}}(u,M_{\gamma}) + g_{\rm rad}(t,u,M_\gamma)
\right]\,,
\end{align}
where the kinematical factor $D(t,u)$ is 
\begin{align}
D(t,u)=\frac{1}{2}m_{\tau}^{2}\left(m_{\tau}^2-t\right)+2m_{\pi}^4-2u\left(m_{\tau}^{2}-t+2m_{\pi}^{2} \right)+2u^{2}\,, 
\end{align}
and the one-loop virtual photon contribution reads~\cite{Cirigliano:2001er} 
\begin{align}
f^{\text{elm}}_{\text{loop}}(u,M_{\gamma})=&\frac{\alpha}{4\pi}\bigg[
\left(u-m_{\pi}^{2}\right)\mathcal{A}(u) +
\left( u-m_{\pi}^{2}-m_{\tau}^{2}
\right)\mathcal{B}(u) \notag \\
&
\qquad +2\left( m_{\pi}^{2}+m_{\tau}^2 -u \right)\mathcal{C}(u,M_{\gamma})
+2\log\frac{m_{\pi}m_{\tau}}{M_{\gamma}^{2}} \bigg]\,,
\end{align}
with 
\begin{align}
\mathcal{A}(u)=
\frac{1}{s}\left(
-\frac{1}{2}\log{ r_{\tau}} +\frac{2-y_{\tau}}{\sqrt{r_{\tau}}}
\frac{x_{\tau}}{1-x_{\tau}^{2}}\log{x_{\tau}}
\right),
\end{align}
\begin{align}
\mathcal{B}(u)=
\frac{1}{s}\left(
\frac{1}{2}\log{ r_{\tau}} +\frac{2r_\tau-y_{\tau}}{\sqrt{r_{\tau}}}
\frac{x_{\tau}}{1-x_{\tau}^{2}}\log{x_{\tau}}
\right),
\end{align}
\begin{align}
\mathcal{C}(u,M_{\gamma})=&
\frac{1}{m_{\tau}m_{\pi}}\frac{x_{\tau}}{1-x_{\tau}^{2}}
\left(
-\frac{1}{2}\log^{2}{x_{\tau}} + 2\log{x_{\tau}}\log{\left(1-x_{\tau}^{2}\right)}
\right. \notag\\
&\left.-\frac{\pi^{2}}{6}+\frac{1}{8}\log^{2}{r_{\tau}}+Li_{2}\left(x_{\tau}^{2}\right)+Li_{2}\left(
1-\frac{x_{\tau}}{\sqrt{r_{\tau}}}
\right)\right.\notag\\
&\left.
+Li_{2}\left(
1-x_{\tau}\sqrt{r_{\tau}}   \right)-\log{x_{\tau}}\log{\frac{M_{\gamma}^{2}}{m_{\tau}m_{\pi}}}
\right),
\end{align}
and  
\begin{align}
r_{\tau}=\frac{m_{\tau}^{2}}{m_{\pi}^{2}}, \quad
y_{\tau}=1+r_{\tau}-\frac{u}{m_{\pi}^{2}}, \quad
x_{\tau}=\frac{1}{2\sqrt{r_{\tau}}}\left(
y_{\tau}-\sqrt{y_{\tau}^{2}-4r_{\tau}}
\right)\,. 
\end{align}
In Eq.~\eqref{eq.dpipiincl}, $g_{\rm rad}$ stands for the real photon corrections calculated from the radiative decay process of $\tauppg$.  
As explicitly demonstrated in detail in Ref.~\cite{Cirigliano:2002pv}, the infrared divergence due to the massless photon in the one-loop diagrams is canceled by the divergence from the bremsstrahlung from the particles $\tau^-$ and $\pi^{-}$ in the soft photon limit, which makes the photon inclusive decay width~\eqref{eq.dpipiincl} infrared finite. In practice, one can retain a tiny finite mass $M_\gamma$ for the photon to separately regulate the infrared divergences arising from the virtual and real photons, i.e., $f^{\text{elm}}_{\text{loop}}$ and $g_{\rm rad}$, when doing numerical calculations.  

To integrate out the $u$ variable in Eq.~\eqref{eq.dpipiincl} with its kinematical allowed upper and lower limits
\begin{align}
u_{\rm max/min}(t)&=\frac{1}{2} \sqrt{1-\frac{4m_{\pi}^{2}}{t}} \left( m_{\tau}^{2} - t
\right)  \pm \frac{1}{2} \left( m_{\tau}^{2}+ 2m_{\pi}^{2}- t\right) \,,
\end{align}
one can obtain the EM correction function $\gem(t)$ for the two-pion tau decay 
\begin{align}\label{eq.dpipiinclt}
\frac{d\Gamma_{\tau_{\pi\pi[\gamma]}}}{dt}=\frac{d\Gamma_{\tau_{\pi\pi}}^{(0)}}{dt}\gem(t)\,.
\end{align}
Since after the integration of $u$ the first term inside the square bracket of Eq.~\eqref{eq.dpipiincl} simply gives the nonradiative decay width~\eqref{eq.nonraddw}, one can conveniently split the virtual ($v$) and real ($r$) photon corrections for $\gem$ as 
\begin{equation}\label{eq.gem}
\gem(t)= 1 + \gem^{(v)}(t) + \gem^{(r)}(t)\,.
\end{equation}
The virtual correction part $\gem^{(v)}(t)$ takes the form 
\begin{align}\label{eq.gemv}
\gem^{(v)}(t)= \frac{12\int^{u_{\rm max}}_{u_{\rm min}} {D(t,u)\, f^{\text{elm}}_{\text{loop}}(u,M_{\gamma})  du } }{m_\tau^6\left(1-\frac{4m_{\pi}^{2}}{t} \right)^{3/2} \left( 1-\frac{t}{m_{\tau}^2} \right)^{2}\left(1+\frac{2t}{m_{\tau}^{2}}\right) } \,. 
\end{align}
The real-photon correction function is similarly given by 
\begin{align}\label{eq.gemrcen}
\gem^{(r)}(t)=  \frac{6\int^{u_{\rm max}}_{u_{\rm min}} { D(t,u)\, g_{\rm rad}(t,u,M_\gamma)   du } }{m_\tau^6\left( 1-\frac{4m_{\pi}^{2}}{t} \right)^{3/2} \left( 1-\frac{t}{m_{\tau}^2} \right)^{2}\left(1+\frac{2t}{m_{\tau}^{2}}
\right)}\,,
\end{align}
where $g_{\rm rad}$ stems from the $\tauppg$ process. Under the leading Low's approximation~\cite{Low:1958sn}, i.e., by only keeping the most singular $1/k^2$ term (with $k$ the photon momentum) and neglecting the terms of $O(k^{n\geq -1})$ for the amplitude squared of the $\tauppg$ process in the soft photon limit $k\to 0$, analytical expressions can be obtained for $g_{\rm rad}(t,u,M_\gamma)$, which has been demonstrated to exactly cancel the infrared divergence of~$f^{\text{elm}}_{\text{loop}}(u,M_{\gamma})$ when taking $M_\gamma\to 0$~\cite{Cirigliano:2002pv}. The sum of the virtual correction $\gem^{(v)}(t)$ and the leading Low (LL) part of the real correction $\gem^{(r,LL)}(t)$, gives a physically meaningful quantity $\gem^{(0)}(t)\equiv\gem^{(v)}(t)+\gem^{(r,LL)}(t)$, which is free of infrared singularity. However, $\gem^{(0)}(t)$ turns out to be insufficient to provide a good approximation in Eq.~\eqref{eq.dpipiinclt}, since the photon inclusive decay width incorporates the dynamics in the whole photon energy range, not only in the low energy, which has been verified in many works~\cite{Cirigliano:2002pv,Flores-Tlalpa:2005msx,Flores-Baez:2006yiq,Miranda:2020wdg}. 

\begin{figure}[htbp]
\centering
\includegraphics[width=0.8\linewidth]{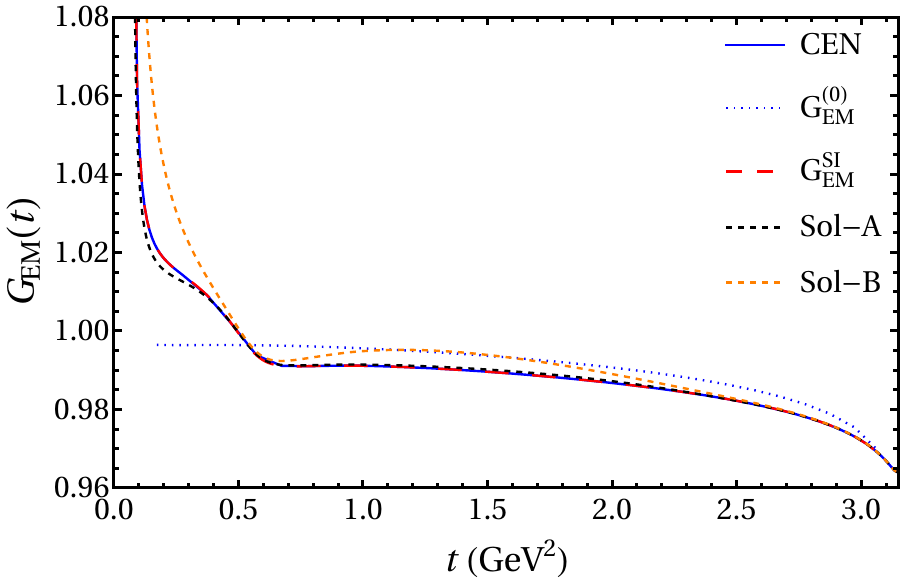}
\caption{
Illustration of different results for EM corrected function $G_{\text{EM}}(t)$: Sol-A (black dashed line, with negative value of $d_4$), Sol-B (orange dashed line, with positive value of $d_4$), $G_{\text{EM}}^{\text{SI}}(t)$ (red long-dashed line, SI term only), the leading Low approximation for $G_{\text{EM}}^{(0)}(t)$ (blue dotted line) and the result with the full amplitude from Ref.~\cite{Cirigliano:2002pv} (blue solid line, labeled as CEN). } \label{fig:gem_all}
\end{figure}

In the present work, instead of using Eq.~\eqref{eq.gemrcen} to obtain the $\gem^{(r)}(t)$, we will directly take the complete $\tauppg$ amplitude and the phase-space receipt from Ref.~\cite{Chen:2022nxm} to calculate the full real radiative correction part of $\gem^{(r)}(t)$. In this case, the latter quantity is given by 
\begin{align}\label{eq.gemrour}
\gem^{(r)}(t)=  \frac{ 1 }{\frac{d\Gamma_{\tau_{\pi\pi}}^{(0)}}{dt}} \frac{\pi^{2}}{32 (2\pi)^{12}m_{\tau}^{2}}\int_{S_{\pi\pi\nu}^-}^{S_{\pi\pi\nu}^+} d S_{\pi\pi\nu} \int_{S_{\nu\gamma}^-}^{S_{\nu\gamma}^+} d S_{\nu\gamma} \int_{S_{\pi\nu}^-}^{S_{\pi\nu}^+} d S_{\pi\nu} \int_{S_{\pi\nu\gamma}^-}^{S_{\pi\nu\gamma}^+}\frac{d S_{\pi\nu\gamma}}{\sqrt{-\Delta_{4}}} \overline{|\mathcal{M}|}^2_{\tau\to\pi\pi\nu_\tau\gamma}\,,
\end{align}
where $\overline{|\mathcal{M}|}^2_{\tau\to\pi\pi\nu_\tau\gamma}$ stands for the amplitude squared after taking the spin sum/average of the initial and final states, $\Delta_4$ is the Gram determinant~\cite{Byckling:1971vca},  $S_{ij\cdots}=(p_i+p_j+\cdots)^2$ and the upper and lower integration limits $S_{ij\cdots}^{\pm}$ are given in the Appendix of Ref.~\cite{Chen:2022nxm}. 
By replacing $\overline{|\mathcal{M}|}^2_{\tau\to\pi\pi\nu_\tau\gamma}$ in Eq.~\eqref{eq.gemrour} with the amplitude from the leading Low expansion and also the full amplitude from Ref.~\cite{Cirigliano:2002pv}, our numerical calculation perfectly reproduces the curves in the latter reference. Moreover, the sum of the virtual correction of Eq.~\eqref{eq.gemv} and the real correction of Eq.~\eqref{eq.gemrour} is verified to be numerically stable by varying the small finite photon mass $M_\gamma$, which reassures the infrared finite feature of the $\gem(t)$ function. The results of the corresponding $\gem(t)$ functions are illustrated in Fig.~\ref{fig:gem_all}.

Next we discuss the results of $\gem(t)$ by implementing the full $\tauppg$ amplitudes mentioned in the previous section into Eq.~\eqref{eq.gemrour} and also combining Eqs.~\eqref{eq.gem} and \eqref{eq.gemv}. 
The curves obtained for the EM corrected function $\gem(t)$ by taking the two different values of $d_4$ in Eqs.~\eqref{eq.d4n} and \eqref{eq.d4p}, labeled as Sol-A and Sol-B in order, together with the results from the leading Low approximation ($\gem^{(0)}$), structure independent (SI) part and the complete amplitude of Ref.~\cite{Cirigliano:2002pv} (labeled as CEN), are shown in Fig.~\ref{fig:gem_all}. We confirm the conclusion of Ref.~\cite{Cirigliano:2002pv} that the leading Low approximation is too rough when calculating the $\gem$ function, see the difference of the curves between $\gem^{(0)}$ and $\gem^{\rm SI}$.  
It turns out that our curve by taking the negative value of $d_4=-0.42$ reconciles with the full result of Ref.~\cite{Cirigliano:2002pv} and the $O(p^4)$ result (with $F_V=2G_V$) of Ref.~\cite{Miranda:2020wdg}, with only slight difference for $t<0.3$~GeV$^2$. Nevertheless, our curve of Sol-B by taking the positive value of $d_4=1.01$ clearly deviates from the latter two results.

\section{Updated evaluation of $a_{\mu}$ from tau data}\label{sec.amu}

In this part, we proceed the evaluation of $a_\mu$ by using the experimental $\taupp$ data, in light of the amended $\gem(t)$ discussed previously. 
At LO of $\alpha$, the HVP contributes to $a_{\mu}$ through~\cite{Gourdin:1969dm}
\begin{align}\label{eq.HVPLO-amu}
a_{\mu}^{\text{HVP},\text{LO}}=\frac{1}{4\pi^{3}}
\int_{4m_\pi^2}^{\infty} dt\, K(t)\,\sigma^{0}_{e^{+}e^{-}\rightarrow \text{hadrons}}\left(t\right),
\end{align}
where the smooth QED kernel function is~\cite{Brodsky:1967sr,Lautrup:1968tdb,Gourdin:1969dm} 
\begin{align}
K(t)=\frac{x^{2}}{2}\left(
2-x^{2}
\right)
+
\frac{\left(
1+x^{2}
\right)\left(
1+x
\right)^{2}}{x^{2}}\left(
\ln\left(1+x\right) -
x+\frac{x^{2}}{2}
\right)+\frac{\left(
1+x
\right)}{\left(
1-x
\right)}x^{2}\ln\left(
x\right)\,, 
\end{align}
with 
\begin{align}
x=\frac{1-\left(1-4m_{\mu}^{2}/t \right)^{1/2}}{1+\left(1-4m_{\mu}^{2}/t \right)^{1/2}}\,. 
\end{align}
The quantity $\sigma^{0}_{e^{+}e^{-}\rightarrow \text{hadrons}}(t)$ corresponds to the undressed hadronic cross sections by excluding the vacuum polarization effects~\cite{Eidelman:1995ny}, and in the case of the two-pion final state it can be written as 
\begin{align}\label{eq.sigpp}
\sigma^{0}_{e^{+}e^{-}\rightarrow\pi^{+}\pi^{-}}=\frac{\pi\alpha^{2}}{3t}\beta_{\pi^{+}\pi^{-}}^3(t) \left|F_{\pi\pi}^{(0)}(t)\right|^2\,,
\end{align}
with  
\begin{equation}
\beta_{\pi^{+}\pi^{-}}(t) = \sqrt{1-\frac{4m_{\pi^+}^2}{t}}\,.
\end{equation}
The conserved vector current indicates that the $\pi^+\pi^-$ EM form factor $F_{\pi\pi}^{(0)}(t)$ in Eq.~\eqref{eq.sigpp} equals to the $\pi^-\pi^0$ vector form factor $F_{\pi\pi}^{(-)}$ in Eq.~\eqref{eq.nonraddw} in the isospin limit. This provides a possible way to calculate the dominant $\pi\pi$ contribution to $a_\mu^{\rm HVP, LO}$ in Eq.~\eqref{eq.HVPLO-amu} by using the experimental data from the $\taupp$ process. 
To reach the subpercent level of precision in this procedure, the isospin breaking (IB) effects must be systematically taken into account. 

\begin{figure}[htbp]
\centering
\includegraphics[width=0.6\linewidth]{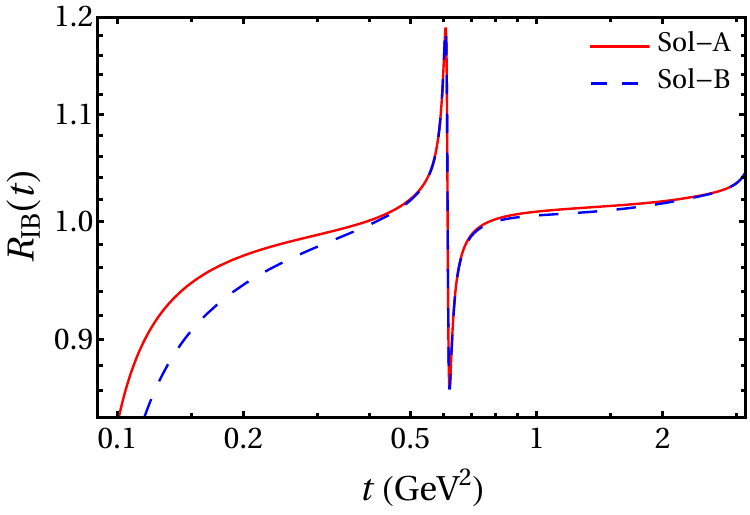}
\caption{
Full IB correction functions $R_{\text{IB}}(t)$ with complete EM corrections $G_{\text{EM}}\left(t\right)$. 
}
\label{fig:RIB_all}
\end{figure}

After including the IB corrections, one can write the $\sigma^{0}_{e^{+}e^{-}\rightarrow\pi^{+}\pi^{-}}$ cross section in terms of the photon inclusive two-pion tau differential decay width~\eqref{eq.dpipiincl} as
\begin{align}\label{eq.sigppdtau}
\sigma^{0}_{e^{+}e^{-}\rightarrow\pi^{+}\pi^{-}}&=\left[
\frac{K_{\sigma}(t)}{K_{\Gamma}(t)}\frac{d\Gamma_{\tau_{\pi\pi[\gamma]}}}{dt}
\right]
\times 
\left[ \frac{R_{\text{IB}}(t)}{S_{\text{EW}}}\right]\,, 
\end{align}
where the kinematical factors are  
\begin{align}
K_{\sigma}(t)=\frac{\pi\alpha^{2}}{3t}\,,\qquad 
K_{\Gamma}(t)=\frac{G_{F}^{2}\left|V_{ud}\right|m_{\tau}^{3}}{384\pi^3} \left(
1-\frac{t}{m_{\tau}^{2}}
\right)^2
\left(
1+2\frac{t}{m_{\tau}^{2}}
\right)\,,
\end{align}
and the IB corrected parts are collected in $R_{\rm IB}(t)$. 
Under the factorization assumption, the various IB corrections can be introduced as products of different pieces~\cite{Cirigliano:2002pv} 
\begin{align}\label{eq.RIB}
R_{\text{IB}}(t)=\frac{\text{FSR}(t)}{G_{\text{EM}}(t)}\frac{\beta_{\pi^{+}\pi^{-}}^{3}(t)}{\beta_{\pi^{-}\pi^{0}}^{3}(t)}\frac{\left|F_{\pi\pi}^{(0)}(t)\right|^{2}}{\left|F_{\pi\pi}^{(-)}(t)\right|^{2}}\,,
\end{align}
where $\text{FSR}(t)$ encloses final-state radiation effect, the ratio of kinematical factors $\beta_{\pi^-\pi^{+,0}}$ caused by the $\pi^{\pm}$ and $\pi^0$ mass difference noticeably deviates from unity near the thresholds, the energy-dependent function $\gem(t)$ encoding the EM corrections to the $\taupp$ process deviates from unity at the level of several percents, as illustrated in Fig.~\ref{fig:gem_all}. Regarding the form factors, it is noted that recently Ref.~\cite{Castro:2024prg} performs a systematic investigation of different form-factor parametrizations, including the Gounaris-Sakurai model \cite{Gounaris:1968mw}, Kuhn-Santamaria model \cite{Kuhn:1990ad}, Guerrero-Pich model~\cite{Guerrero:1997ku} and the dispersive form~\cite{Colangelo:2018mtw}. 
In order to make a close comparison with Ref.~\cite{Cirigliano:2002pv}, we will stick to the same form factors of Ref.~\cite{Guerrero:1997ku} used in the former reference 
\begin{align}
F_{\pi\pi}^{(0)}(t)&=M_{\rho^{0}}^{2}D_{\rho^{0}}^{-1}\left(t\right)\left[
\exp\left(
2\widetilde{H}_{\pi^{+}\pi^{-}}\left(t\right)+
\widetilde{H}_{K^{+}K^{-}}\left(t\right)
\right)-
\frac{\theta_{\rho\omega}}{3M_{\rho}^{2}} 
\frac{t}{M_{\omega}^{2}-t-iM_{\omega}\Gamma_{\omega}}
\right], \\
F_{\pi\pi}^{(-)}(t)&= M_{\rho^{+}}^{2}D_{\rho^{+}}^{-1}\left(t\right) 
\exp\left(
2\widetilde{H}_{\pi^{-}\pi^{0}}\left(t\right)
+\widetilde{H}_{K^{-}K^{0}}\left(t\right)
\right)
+f_{\text{local}}^{\text{elm}}+... \quad \,,
\end{align}
where the loop function $\widetilde{H}_{P,Q}\left(t\right)$ and local EM correction terms $f_{\text{local}}^{\text{elm}}$ have been explicitly provided in Ref.~\cite{Cirigliano:2002pv}. The values of the parameters $M_\rho$, $\Gamma_\rho$, $\theta_{\rho\omega}$, appearing in the above expression will be taken the same as those in the latter reference. The FSR($t$) function is elaborated in detail in Refs.~\cite{Melnikov:2001uw,Jegerlehner:2017gek} and we directly take the results from these references. 
The result for the full IB correction function $R_{\rm IB}(t)$ is shown in Fig.~\ref{fig:RIB_all}, which demonstrates that $R_{\rm IB}(t)$ derived from different parameter sets exhibit excellent agreement in the energy region of $t>0.5$~GeV$^2$. The deviations from different sets of $d_4$ parameters  manifest in the low energy region of $t<0.4$~GeV$^2$.

\begin{table}[!ht]
\centering
\begin{tabular}{l| c c c c c c} 
\hline
\hline
\multicolumn{7}{c}{$\Delta a_{\mu}^{\text{HVP},\text{LO}}[\pi\pi]$ from $\gem(t)$}\\
\hline
$[t_{min},t_{max}]$ & Sol-A & Sol-B & CEN &\multicolumn{2}{c}{MR[$\mathcal{O}(p^{4})$]}&MR[$\mathcal{O}(p^{6})$]  \\ [0.5ex] 
\hline
$\left[ 4m_{\pi}^{2},1\,\text{GeV}^{2}\right] $ 
& $-$ 7.1 & $-$44.9 & $-$10.6 & $-$10.4 & $-$15.9 & $-$63.2$\pm$16.5 
\\ 
$\left[ 4m_{\pi}^{2},2\,\text{GeV}^{2}\right] $ 
& $-$ 6.4 & $-$44.5  & $-$9.8 & $-$9.6 & $-$15.2  & $-$58.1$\pm$12.2 
\\ 
$\left[ 4m_{\pi}^{2},3\,\text{GeV}^{2}\right] $ 
& $-$ 6.3  & $-$44.4 & $-$9.7 & $-$9.5 & $-$15.1 & $-$67.8$\pm$17.5 
\\
$\left[ 4m_{\pi}^{2},m_{\tau}^{2}\right] $ 
& $-$ 6.3 & $-$44.4  & $-$9.7 & $-$9.5 & $-$15.1 & $-$64.9$\pm$13.4 
\\ 
\hline
\hline
\end{tabular}
\caption{
The contributions of $G_{\text{EM}}\left(t\right)$ to the relative shifts of $\Delta a_{\mu}^{\text{HVP},\text{LO}}$ (in units of $10^{-11}$) as defined in \eqref{eq.deltamu}. The numbers in different rows illustrate the results by taking the indicated energy intervals shown in the first column to evaluate the integral in Eq.~\eqref{eq.deltamu}. The second and third columns represent the results of the parameter sets of Sol-A and Sol-B in our study, respectively. The fourth column shows the results of Ref.~\cite{Cirigliano:2002pv}, and the fifth to seventh columns present the outcomes from $F_{V}=\sqrt{2}F$ and $F_{V}=\sqrt{3}F$ at $\mathcal{O}(p^{4})$ , as well as those from $\mathcal{O}(p^{6})$ in Ref.~\cite{Miranda:2020wdg}, in order. }
\label{tab:Gem}
\end{table}

\begin{table}[htbp]
\centering
\begin{tabular}{c| c c c c c}
\hline
\hline
\multicolumn{6}{c}{$\Delta a_{\mu}^{\text{HVP},\text{LO}}[\pi\pi]$ from various IB sources} \\  
\hline
Parameters&$t_{max}/\text{GeV}^2$ & $S_{\text{EW}}$ & $\beta_{\pi^+\pi^-}^3/\beta_{\pi^-\pi^0}^3$ & $\big|F_{\pi\pi}^{(0)}/F_{\pi\pi}^{(-)}\big|^2$ &FSR  \\ [0.5ex] 
\hline
\multirow{4}{3em}{Sol-A}
&$ 1.0 $ & $-$112.7 & $-$74.1 & 76.8 & 42.7  \\
&$ 2.0 $ & $-$114.6 & $-$74.3 & 76.7 & 43.3  \\
&$ 3.0 $ & $-$114.7 & $-$74.3 & 76.7 & 43.3  \\
&$m_{\tau}^{2}$ & $-$114.7& $-$74.3 & 76.7 & 43.3  \\
\hline
\multirow{4}{3em}{Sol-B}
&$ 1.0 $ & $-$113.6 & $-$76.9 & 77.1 & 43.2  \\ 
&$ 2.0 $ & $-$115.4 & $-$77.1 & 77.0 & 43.8  \\ 
&$ 3.0 $ & $-$115.5 & $-$77.1 & 77.0 & 43.8  \\
&$m_{\tau}^{2}$ & $-$115.5 & $-$77.1 & 77.0 & 43.8  \\ 
\hline
\hline
\end{tabular}
\caption{
Contributions to the relative shifts of $\Delta a_{\mu}^{\text{HVP},\text{LO}}[\pi\pi]$ as defined in \eqref{eq.deltamu} from various IB sources (in units of $10^{-11}$) by taking different values of $t_{max}$(in units of $\text{GeV}^{2}$) . }
\label{tab:other-term}
\end{table}

\begin{table}[htbp]
       \centering
       \begin{tabular}{c| c c c c }
       \hline
       \hline
       Parameters& Experiments  & $a_{\mu}^{\text{HVP,LO}|_{\pi\pi,\tau \text{data}}}$  
       \\
       \hline
       \multirow{4}{3em}{Sol-A}
       &Belle& $516.7 \pm2.1 \pm 7.9  
        \pm 2.2$ 
        \\
       &ALEPH & $513.3\pm4.3 \pm 2.8  
       \pm 2.1 $ & &\\
       &CLEO & $516.9 \pm3.2 \pm 8.8  
       \pm 2.2$ & &\\
       &OPAL & $527.2  \pm 9.8 \pm 6.8 
       \pm 2.1
       $& &\\
       \hline
       \multirow{4}{3em}{Sol-B}
       &Belle & $513.4 \pm2.0 \pm 7.9   
       \pm 2.2$
       \\
       &ALEPH &$510.2 \pm4.2 \pm 2.8   
       \pm 2.1$  & &\\
       &CLEO &$513.7 \pm3.2 \pm 8.8   
       \pm 2.2
       $  & &\\
       &OPAL & $523.5 \pm9.5 \pm 6.8 
       \pm 2.1 
       $ & &\\
       \hline
        \hline
       \end{tabular}
       \caption{ The IB-corrected $\pi\pi$ contribution to $a_{\mu}^{\text{HVP},\text{LO}}$ in units of $10^{-10}$ by using various experimental $\taupp$ data. 
For the uncertainties in the third column, the first error is from the uncertainties of the experimental invariant-mass spectra, the second error arises from the branchings ratios of $\mathcal{B}_{\tau_{\pi\pi}}$ and $\mathcal{B}_{\tau_e}$, and the third error comes from the IB correction function $R_{\rm IB}$. 
       }
       \label{tab:amu}
       \end{table}

Next, we calculate the relative shifts of $a_\mu$ caused by the IB corrections in the $\pi\pi$ channel via the following formula
\begin{align}\label{eq.deltamu}
\Delta a_{\mu}^{\text{HVP,LO}}=\frac{1}{4\pi^3}\int_{4m_{\pi}^{2}}^{t_{max}} dt
\,K(t)\left[
\frac{K_{\sigma}(t)}{K_{\Gamma}(t)}\frac{d\Gamma_{\tau_{\pi\pi[\gamma]}}}{dt}
\right]
\times 
\left( \frac{R_{\text{IB}}(t)}{S_{\text{EW}}}-1\right) \,,
\end{align} 
where the photon inclusive two-pion tau differential decay width is given in Eq.~\eqref{eq.dpipiinclt}. Each source of the IB corrections is separately analyzed and the corresponding results are summarized in Table~\ref{tab:Gem} for $\gem(t)$ and Table~\ref{tab:other-term} for others, such as $\beta_{\pi^-\pi^{+}}^3/\beta_{\pi^-\pi^{0}}^3$, $S_{\rm EW}$, $|F_{\pi\pi}^{(0)}/F_{\pi\pi}^{(-)}|^2$ and the FSR effects. 
As shown in Table~\ref{tab:Gem}, the magnitude of $\Delta a_{\mu}^{\text{HVP,LO}}$ tends to decrease and stabilize when $t>2$~GeV$^2$, indicating that the dominant contributions arise from the low-energy regions.
Through the comparison of the two parameter sets, we observe that the results of $\gem(t)$ in the case of Sol-A with negative value of $d_4=-0.42$, is rather different from those of Sol-B with positive value of $d_4=1.01$. This indicates that the effect from $\gem(t)$ is rather sensitive to the parameter $d_4$, especially about its positive/negative sign. As discussed previously, the inclusion of the scalar resonance effects in $\omega\to\pi^0\pi^0\gamma$ decay can moderately affect the determination of $d_4$, which in turn will modify $G_{\rm EM}(t)$ and $\Delta a_{\mu}^{\text{HVP,LO}}$. It is verified that the latter quantity will be shifted around 5\% in the case of positive solution of $d_4$ (Sol-B) by including/excluding the scalar effects in $\omega\to\pi^0\pi^0\gamma$, while in the case with negative solution of $d_4$ (Sol-A) $\Delta a_{\mu}^{\text{HVP,LO}}$ is barely changed.
In fact, the sensitivity of the sign of $d_4$ in the $\tauppg$ process has been pointed out in Ref.~\cite{Chen:2022nxm}, where distinguished curves are predicted to the invariant-mass and T-odd distributions for the different parameters sets. A future experimental measurement on these distributions can be definitely helpful to pin down the value of $d_4$. 

A brief comparison among the results of ours based on the amplitude of Ref.~\cite{Chen:2022nxm} and those in Refs.~\cite{Cirigliano:2002pv,Miranda:2020wdg} is in order. The key difference of the three references lies in different treatments of the structure-dependent form factors in the $\tau\to\pi\pi\gamma\nu_\tau$ decays that will in turn affect the long-distance EM correction function $G_{\rm EM}(t)$. The $O(p^4)$ amplitude of Ref.~\cite{Miranda:2020wdg} coincides with that of Ref.~\cite{Cirigliano:2002pv}, which employs the minimal even-intrinsic-parity $R \chi T$ operators in the calculation of the $\tau\to\pi\pi\gamma\nu_\tau$ process. While, the  $O(p^6)$ amplitude of Ref.~\cite{Miranda:2020wdg} extends the result of Ref.~\cite{Cirigliano:2002pv} by including more complete resonance operators with both even- and odd-intrinsic-parity types, though many of the resonance couplings are unknown and roughly estimated by simple dimensional analysis with rather large uncertainties. The current work uses an alternative resonance operator base from Ref.~\cite{Ruiz-Femenia:2003jdx} to include the relevant odd-intrinsic parity interaction, which improves the description of  Ref.~\cite{Cirigliano:2002pv} relying on the minimal $R\chi T$ Lagrangian and also provides a complementary study to Ref.~\cite{Miranda:2020wdg}, since we find a way to fix all the unknown resonance couplings, though with two solutions of $d_4$.
Interestingly, the results of $\gem$ from the Sol-A scenario are similar with those from Ref.~\cite{Cirigliano:2002pv} (labeled as CEN) and one of the $\mathcal{O}(p^4)$ cases of Ref.~\cite{Miranda:2020wdg} (denoted as MR[$\mathcal{O}(p^4)$]), while the results of $\gem$ from Sol-B are close to the $\mathcal{O}(p^6)$ case of the former reference (denoted as MR[$\mathcal{O}(p^6)$]). 
In Table~\ref{tab:other-term}, we also present the individual contributions to the relative shifts of $\Delta a_{\mu}$ from the short-distance electroweak correction $S_{\text{EW}}$, the phase space factor $\beta_{\pi^{+}\pi^{-}}^{3}/\beta_{\pi^{-}\pi^{0}}^{3}$, the ratio of form factors $\big|F_{\pi\pi}^{(0)}(t)/F_{\pi\pi}^{(-)}(t)\big|^{2}$ and the FSR correction.

Alternatively, with the established IB correction function $R_{\rm IB}(t)$ one can also extract the $e^{+}e^{-}\rightarrow\pi^{+}\pi^{-}$ cross section using Eq.~\eqref{eq.sigppdtau} from the experimental $\pi\pi$ invariant-mass distribution of $\tau$ decay, which can be conveniently recast as 
\begin{align}\label{eq.sigppdtaupp}
\sigma^{0}_{e^+e^-\to\pi^+\pi^-}(t) =\frac{2\pi\alpha^{2}m_{\tau}^{2}}{3| V_{ud}|^{2}t}\frac{1}{\left(
1-\frac{t}{m_{\tau}^2}
\right)^{2}\left(
1+\frac{2t}{m_{\tau}^{2}}
\right)}\frac{\mathcal{B}_{\tau_{\pi\pi}}}{\mathcal{B}_{\tau_e}} \frac{1}{N_{\tau_{\pi\pi}}}\frac{dN_{\tau_{\pi\pi}}}{dt}\frac{R_{\text{IB}}\left(t\right)}{S_{\text{EW}}}\,,
\end{align}
where $dN_{\tau_{\pi\pi}}/(N_{\tau_{\pi\pi}}\,dt)$ is the normalized $\pi\pi$ invariant-mass distribution from the $\tau\to\pi\pi\nu_\tau$ process, and 
$\mathcal{B}_{\tau_{\pi\pi}}$ and $\mathcal{B}_{\tau_e}$ denote the branching ratios of $\tau\to\pi\pi\nu_\tau$ and $\tau \to e\nu_{\tau}\bar{\nu}_{e}$, respectively. 
Given the scarcity of experimental data near the $\pi\pi$ threshold in $\tau$ decays, we will use the dispersive results from Ref. \cite{Gonzalez-Solis:2019iod} as input, instead of using the actual experimental data, in the  energy range of $[4m_{\pi}^{2} , 0.1296~{\rm GeV}^2]$, as also done in Refs.~\cite{Davier:2010fmf,Miranda:2020wdg}. For the higher energy range $[0.1296~{\rm GeV}^2, m_{\tau}^{2}]$, we will directly take the experimental data from Belle~\cite{Belle:2008xpe}, ALEPH~\cite{Davier:2013sfa}, CLEO~\cite{CLEO:1999dln} and OPAL~\cite{OPAL:1998rrm}. 
By substituting the cross section of Eq.~\eqref{eq.sigppdtaupp} into Eq.~\eqref{eq.HVPLO-amu}, we can calculate the LO HVP contribution to $a_\mu$ from the $\pi\pi$ channel based on the tau data, and the corresponding numerical results are collected in Table~\ref{tab:amu}. It is noted that the available correlations among the different data points from each experiment have been taken into account in our uncertainty analyses.

An argument based on simple dimensional analysis~\cite{Miranda:2020wdg} hints somewhat small magnitude of $d_4$, so we will set the results by taking the parameter Sol-A with $d_4=-0.42$ as our baseline in this work. Furthermore, we consider the deviation from the Sol-B parameter set as an additional theoretical systematic uncertainty. The corresponding baseline results from Sol-A by using different experimental $\taupp$ data are 
\begin{align}\label{eq.amubelle}
a_{\mu}^{\text{HVP,LO}|_{\pi\pi,\tau \text{data, Belle}}}=&\,
516.7\pm 2.1_{\text{Spec}} \pm 7.9_{\text{BR}}\pm 2.2_{\text{IB}}\pm3.3_{\text{Sys}}\,, \\
a_{\mu}^{\text{HVP,LO}|_{\pi\pi,\tau \text{data, ALEPH}}}=&\,
513.3\pm4.3_{\text{Spec}}  \pm 2.8_{\text{BR}}\pm 2.1_{\text{IB}}\pm 3.2_{\text{Sys}}\,,\\
a_{\mu}^{\text{HVP,LO}|_{\pi\pi,\tau \text{data, CLEO}}}=&\, 516.9 \pm3.2_{\text{Spec}} \pm 8.9_{\text{BR}}\pm 2.2_{\text{IB}}\pm 3.3_{\text{Sys}}\,,\\
a_{\mu}^{\text{HVP,LO}|_{\pi\pi,\tau \text{data, OPAL}}}=&\,
527.2  \pm9.8_{\text{Spec}} \pm 6.8_{\text{BR}}\pm 2.1_{\text{IB}}\pm 3.7_{\text{Sys}}\,, \label{eq.amuopal}
\end{align}
where the second (labeled as Spec) and third (labeled as BR) entries in the right-hand side denote the uncertainties  caused by experimental data of the spectral $\pi\pi$ distribution and the branching ratios of $\mathcal{B}_{\tau_{\pi\pi}}$ and $\mathcal{B}_{\tau_e}$ (BR), the fourth entry corresponds to the uncertainty from the IB correction,  and the last entry represent the theoretical systematical uncertainties estimated by taking the differences of the central values from Sol-A and Sol-B.

After performing the average of the central values in Eqs.~\eqref{eq.amubelle}-\eqref{eq.amuopal} weighted by the experimental error bars of the invariant-mass spectra and branching ratios, we obtain the contribution from the $\pi\pi$ channel to $a_{\mu}^{\text{HVP,LO}}$ by using the tau data as 
\begin{align}
10^{10}\cdot a_{\mu}^{\text{HVP,LO}|_{\pi\pi,\tau \text{data}}}=516.0 \pm 2.9_{\text{Spec}+\text{BR}}\pm 4.0_{\text{IB}+\text{Sys}}\,,
\end{align}
which can be compared to the value of $517.2\pm 2.8_{\rm Exp}\pm 5.1_{\rm Th}$ estimated in WP25~\cite{Aliberti:2025beg}. 
By including the results from other hadronic channels in $e^{+}e^{-}$ annihilation from Ref.~\cite{Davier:2019can}, the full LO HVP contribution to $a_\mu$ reads 
\begin{align}\label{eq.amuhvplo}
10^{10}\cdot a_{\mu}^{\text{HVP,LO}|_{\tau \text{data}}}=702.1 \pm 2.9_{\text{Spec}+\text{BR}}\pm 4.0_{\text{IB}+\text{Sys}} \pm 2.1_{\text{Others}}\,,
\end{align}
where the last entry corresponds to the uncertainties caused by the channels with higher thresholds from the $e^{+}e^{-}$ data and the perturbative QCD contributions~\cite{Miranda:2020wdg}. By adding the error bars quadratically in Eq.~\eqref{eq.amuhvplo}, we acquire $10^{10}\cdot a_{\mu}^{\text{HVP,LO}|_{\tau \text{data}}}=702.1\pm 5.4$. By further including the revised contributions from electromagnetic, weak and other hadronic interactions given in WP25~\cite{Aliberti:2025beg}, such as the hadronic light-by-light corrections, the deviation between the Standard Model prediction and combined BNL+FNAL experimental result \cite{Muong-2:2023cdq,Aliberti:2025beg} is 
\begin{align}
\Delta a_{\mu} \equiv  a_{\mu}^{\text{exp}} -a_{\mu}^{\text{SM}}=(14.9 \pm 5.6) \times 10^{-10}\,,
\end{align}
leading to a 2.7$\sigma$ discrepancy. Therefore our study confirms that although the SM prediction to $a_\mu$ based on the $\tau\to\pi\pi\nu_\tau$ data is somewhat below the value by using the lattice QCD result as advocated in WP25~\cite{Aliberti:2025beg}, such $a_\mu$ value based on tau data is clearly larger than that of WP20 based on the experimental $e^+e^-$ annihilation data without the CMD-3 measurement.

\section{Summary and conclusions}\label{sec.concl}

We have revisited the computation of the muon anomalous magnetic moment $a_\mu$ by using the experimental $\tau\to\pi\pi\nu_\tau$ data to assess the leading-order hadronic vacuum polarization from the $\pi\pi$ channel. The real-photon electromagnetic correction to the $\tau\to\pi\pi\nu_\tau$ process has been calculated based on the $\tau\to\pi\pi\gamma\nu_\tau$ amplitude, which is computed by taking the minimal resonance chiral theory with even parity operators and also the anomalous resonance operators with the $VVP$ and $VJP$ types. After imposing the high energy constraints, the only unknown resonance coupling of the $VVP$ type, namely $d_4$, is newly determined through the $\omega\to\pi^0\pi^0\gamma$ process by including both the vector and scalar resonances. The negative solution of $d_4=-0.42$ gives rather a similar long-distance electromagnetic correction function $\gem(t)$ as the previous determination in Ref.~\cite{Cirigliano:2002pv} and also one set of the $O(p^4)$ results in Ref.~\cite{Miranda:2020wdg}. While, the positive solution of $d_4=1.01$ obviously deviates from the former two results, but gives compatible result as the $O(p^6)$ case in the latter reference. We therefore take the negative solution of $d_4$ as the baseline in this work and consider the difference caused by the two solutions of $d_4$ as additional theoretical systematical uncertainty. 

After combining the isospin breaking effects from other sources, we acquire the full set of isospin breaking correction function $R_{\rm{IB}}(t)$, which is further exploited to calculate $a_\mu^{\rm HVP,LO|_{\pi\pi}}$ by using the experimental $\tau\to\pi\pi\nu_\tau$ data from Belle, ALEPH, CLEO and OPAL. Our result agrees with the conservative estimation provided in the muon $g-2$ White Paper 2025 within the uncertainties. The resulting deviation of $a_\mu$ between the current estimation and the updated world average in Ref.~\cite{Muong-2:2025xyk} lies at the level of 2.7~$\sigma$.

\section*{Acknowledgements}

We would like to thank Changzheng Yuan and Zhiqing Zhang for providing and elaborating the experimental tau data. This work is supported in part by National Natural Science Foundation of China (NSFC) of China under Grants No.~12475078, No.~12150013, No.~11975090, and also partially supported by the Science Foundation of Hebei Normal University with Contract No.~L2023B09. 


\bibliography{amu}

\begin{thebibliography}{66}%
\makeatletter
\providecommand \@ifxundefined [1]{%
 \@ifx{#1\undefined}
}%
\providecommand \@ifnum [1]{%
 \ifnum #1\expandafter \@firstoftwo
 \else \expandafter \@secondoftwo
 \fi
}%
\providecommand \@ifx [1]{%
 \ifx #1\expandafter \@firstoftwo
 \else \expandafter \@secondoftwo
 \fi
}%
\providecommand \natexlab [1]{#1}%
\providecommand \enquote  [1]{``#1''}%
\providecommand \bibnamefont  [1]{#1}%
\providecommand \bibfnamefont [1]{#1}%
\providecommand \citenamefont [1]{#1}%
\providecommand \href@noop [0]{\@secondoftwo}%
\providecommand \href [0]{\begingroup \@sanitize@url \@href}%
\providecommand \@href[1]{\@@startlink{#1}\@@href}%
\providecommand \@@href[1]{\endgroup#1\@@endlink}%
\providecommand \@sanitize@url [0]{\catcode `\\12\catcode `\$12\catcode
  `\&12\catcode `\#12\catcode `\^12\catcode `\_12\catcode `\%12\relax}%
\providecommand \@@startlink[1]{}%
\providecommand \@@endlink[0]{}%
\providecommand \url  [0]{\begingroup\@sanitize@url \@url }%
\providecommand \@url [1]{\endgroup\@href {#1}{\urlprefix }}%
\providecommand \urlprefix  [0]{URL }%
\providecommand \Eprint [0]{\href }%
\providecommand \doibase [0]{https://doi.org/}%
\providecommand \selectlanguage [0]{\@gobble}%
\providecommand \bibinfo  [0]{\@secondoftwo}%
\providecommand \bibfield  [0]{\@secondoftwo}%
\providecommand \translation [1]{[#1]}%
\providecommand \BibitemOpen [0]{}%
\providecommand \bibitemStop [0]{}%
\providecommand \bibitemNoStop [0]{.\EOS\space}%
\providecommand \EOS [0]{\spacefactor3000\relax}%
\providecommand \BibitemShut  [1]{\csname bibitem#1\endcsname}%
\let\auto@bib@innerbib\@empty
\bibitem [{\citenamefont {Abi}\ \emph {et~al.}(2021)\citenamefont {Abi} \emph
  {et~al.}}]{Muong-2:2021ojo}%
  \BibitemOpen
  \bibfield  {author} {\bibinfo {author} {\bibfnamefont {B.}~\bibnamefont
  {Abi}} \emph {et~al.} (\bibinfo {collaboration} {Muon g-2}),\ }\href
  {https://doi.org/10.1103/PhysRevLett.126.141801} {\bibfield  {journal}
  {\bibinfo  {journal} {Phys. Rev. Lett.}\ }\textbf {\bibinfo {volume} {126}},\
  \bibinfo {pages} {141801} (\bibinfo {year} {2021})},\ \Eprint
  {https://arxiv.org/abs/2104.03281} {arXiv:2104.03281 [hep-ex]} \BibitemShut
  {NoStop}%
\bibitem [{\citenamefont {Aguillard}\ \emph {et~al.}(2023)\citenamefont
  {Aguillard} \emph {et~al.}}]{Muong-2:2023cdq}%
  \BibitemOpen
  \bibfield  {author} {\bibinfo {author} {\bibfnamefont {D.~P.}\ \bibnamefont
  {Aguillard}} \emph {et~al.} (\bibinfo {collaboration} {Muon g-2}),\ }\href
  {https://doi.org/10.1103/PhysRevLett.131.161802} {\bibfield  {journal}
  {\bibinfo  {journal} {Phys. Rev. Lett.}\ }\textbf {\bibinfo {volume} {131}},\
  \bibinfo {pages} {161802} (\bibinfo {year} {2023})},\ \Eprint
  {https://arxiv.org/abs/2308.06230} {arXiv:2308.06230 [hep-ex]} \BibitemShut
  {NoStop}%
\bibitem [{\citenamefont {Aguillard}\ \emph {et~al.}(2025)\citenamefont
  {Aguillard} \emph {et~al.}}]{Muong-2:2025xyk}%
  \BibitemOpen
  \bibfield  {author} {\bibinfo {author} {\bibfnamefont {D.~P.}\ \bibnamefont
  {Aguillard}} \emph {et~al.} (\bibinfo {collaboration} {Muon g-2}),\
  }\href@noop {} {\  (\bibinfo {year} {2025})},\ \Eprint
  {https://arxiv.org/abs/2506.03069} {arXiv:2506.03069 [hep-ex]} \BibitemShut
  {NoStop}%
\bibitem [{\citenamefont {Aliberti}\ \emph {et~al.}(2025)\citenamefont
  {Aliberti} \emph {et~al.}}]{Aliberti:2025beg}%
  \BibitemOpen
  \bibfield  {author} {\bibinfo {author} {\bibfnamefont {R.}~\bibnamefont
  {Aliberti}} \emph {et~al.},\ }\href@noop {} {\  (\bibinfo {year} {2025})},\
  \Eprint {https://arxiv.org/abs/2505.21476} {arXiv:2505.21476 [hep-ph]}
  \BibitemShut {NoStop}%
\bibitem [{\citenamefont {Aoyama}\ \emph {et~al.}(2020)\citenamefont {Aoyama}
  \emph {et~al.}}]{Aoyama:2020ynm}%
  \BibitemOpen
  \bibfield  {author} {\bibinfo {author} {\bibfnamefont {T.}~\bibnamefont
  {Aoyama}} \emph {et~al.},\ }\href
  {https://doi.org/10.1016/j.physrep.2020.07.006} {\bibfield  {journal}
  {\bibinfo  {journal} {Phys. Rept.}\ }\textbf {\bibinfo {volume} {887}},\
  \bibinfo {pages} {1} (\bibinfo {year} {2020})},\ \Eprint
  {https://arxiv.org/abs/2006.04822} {arXiv:2006.04822 [hep-ph]} \BibitemShut
  {NoStop}%
\bibitem [{\citenamefont {Ignatov}\ \emph
  {et~al.}(2024{\natexlab{a}})\citenamefont {Ignatov} \emph
  {et~al.}}]{CMD-3:2023rfe}%
  \BibitemOpen
  \bibfield  {author} {\bibinfo {author} {\bibfnamefont {F.~V.}\ \bibnamefont
  {Ignatov}} \emph {et~al.} (\bibinfo {collaboration} {CMD-3}),\ }\href
  {https://doi.org/10.1103/PhysRevLett.132.231903} {\bibfield  {journal}
  {\bibinfo  {journal} {Phys. Rev. Lett.}\ }\textbf {\bibinfo {volume} {132}},\
  \bibinfo {pages} {231903} (\bibinfo {year} {2024}{\natexlab{a}})},\ \Eprint
  {https://arxiv.org/abs/2309.12910} {arXiv:2309.12910 [hep-ex]} \BibitemShut
  {NoStop}%
\bibitem [{\citenamefont {Ignatov}\ \emph
  {et~al.}(2024{\natexlab{b}})\citenamefont {Ignatov} \emph
  {et~al.}}]{CMD-3:2023alj}%
  \BibitemOpen
  \bibfield  {author} {\bibinfo {author} {\bibfnamefont {F.~V.}\ \bibnamefont
  {Ignatov}} \emph {et~al.} (\bibinfo {collaboration} {CMD-3}),\ }\href
  {https://doi.org/10.1103/PhysRevD.109.112002} {\bibfield  {journal} {\bibinfo
   {journal} {Phys. Rev. D}\ }\textbf {\bibinfo {volume} {109}},\ \bibinfo
  {pages} {112002} (\bibinfo {year} {2024}{\natexlab{b}})},\ \Eprint
  {https://arxiv.org/abs/2302.08834} {arXiv:2302.08834 [hep-ex]} \BibitemShut
  {NoStop}%
\bibitem [{\citenamefont {Alemany}\ \emph {et~al.}(1998)\citenamefont
  {Alemany}, \citenamefont {Davier},\ and\ \citenamefont
  {Hocker}}]{Alemany:1997tn}%
  \BibitemOpen
  \bibfield  {author} {\bibinfo {author} {\bibfnamefont {R.}~\bibnamefont
  {Alemany}}, \bibinfo {author} {\bibfnamefont {M.}~\bibnamefont {Davier}},\
  and\ \bibinfo {author} {\bibfnamefont {A.}~\bibnamefont {Hocker}},\ }\href
  {https://doi.org/10.1007/s100520050127} {\bibfield  {journal} {\bibinfo
  {journal} {Eur. Phys. J. C}\ }\textbf {\bibinfo {volume} {2}},\ \bibinfo
  {pages} {123} (\bibinfo {year} {1998})},\ \Eprint
  {https://arxiv.org/abs/hep-ph/9703220} {arXiv:hep-ph/9703220} \BibitemShut
  {NoStop}%
\bibitem [{\citenamefont {Cirigliano}\ \emph {et~al.}(2002)\citenamefont
  {Cirigliano}, \citenamefont {Ecker},\ and\ \citenamefont
  {Neufeld}}]{Cirigliano:2002pv}%
  \BibitemOpen
  \bibfield  {author} {\bibinfo {author} {\bibfnamefont {V.}~\bibnamefont
  {Cirigliano}}, \bibinfo {author} {\bibfnamefont {G.}~\bibnamefont {Ecker}},\
  and\ \bibinfo {author} {\bibfnamefont {H.}~\bibnamefont {Neufeld}},\ }\href
  {https://doi.org/10.1088/1126-6708/2002/08/002} {\bibfield  {journal}
  {\bibinfo  {journal} {JHEP}\ }\textbf {\bibinfo {volume} {08}},\ \bibinfo
  {pages} {002}},\ \Eprint {https://arxiv.org/abs/hep-ph/0207310}
  {arXiv:hep-ph/0207310} \BibitemShut {NoStop}%
\bibitem [{\citenamefont {Cirigliano}\ \emph {et~al.}(2001)\citenamefont
  {Cirigliano}, \citenamefont {Ecker},\ and\ \citenamefont
  {Neufeld}}]{Cirigliano:2001er}%
  \BibitemOpen
  \bibfield  {author} {\bibinfo {author} {\bibfnamefont {V.}~\bibnamefont
  {Cirigliano}}, \bibinfo {author} {\bibfnamefont {G.}~\bibnamefont {Ecker}},\
  and\ \bibinfo {author} {\bibfnamefont {H.}~\bibnamefont {Neufeld}},\ }\href
  {https://doi.org/10.1016/S0370-2693(01)00764-X} {\bibfield  {journal}
  {\bibinfo  {journal} {Phys. Lett. B}\ }\textbf {\bibinfo {volume} {513}},\
  \bibinfo {pages} {361} (\bibinfo {year} {2001})},\ \Eprint
  {https://arxiv.org/abs/hep-ph/0104267} {arXiv:hep-ph/0104267} \BibitemShut
  {NoStop}%
\bibitem [{\citenamefont {Ecker}\ \emph
  {et~al.}(1989{\natexlab{a}})\citenamefont {Ecker}, \citenamefont {Gasser},
  \citenamefont {Pich},\ and\ \citenamefont {de~Rafael}}]{Ecker:1988te}%
  \BibitemOpen
  \bibfield  {author} {\bibinfo {author} {\bibfnamefont {G.}~\bibnamefont
  {Ecker}}, \bibinfo {author} {\bibfnamefont {J.}~\bibnamefont {Gasser}},
  \bibinfo {author} {\bibfnamefont {A.}~\bibnamefont {Pich}},\ and\ \bibinfo
  {author} {\bibfnamefont {E.}~\bibnamefont {de~Rafael}},\ }\href
  {https://doi.org/10.1016/0550-3213(89)90346-5} {\bibfield  {journal}
  {\bibinfo  {journal} {Nucl. Phys. B}\ }\textbf {\bibinfo {volume} {321}},\
  \bibinfo {pages} {311} (\bibinfo {year} {1989}{\natexlab{a}})}\BibitemShut
  {NoStop}%
\bibitem [{\citenamefont {Flores-Tlalpa}\ \emph {et~al.}(2005)\citenamefont
  {Flores-Tlalpa}, \citenamefont {Lopez~Castro},\ and\ \citenamefont
  {Sanchez~Toledo}}]{Flores-Tlalpa:2005msx}%
  \BibitemOpen
  \bibfield  {author} {\bibinfo {author} {\bibfnamefont {A.}~\bibnamefont
  {Flores-Tlalpa}}, \bibinfo {author} {\bibfnamefont {G.}~\bibnamefont
  {Lopez~Castro}},\ and\ \bibinfo {author} {\bibfnamefont {G.}~\bibnamefont
  {Sanchez~Toledo}},\ }\href {https://doi.org/10.1103/PhysRevD.72.113003}
  {\bibfield  {journal} {\bibinfo  {journal} {Phys. Rev. D}\ }\textbf {\bibinfo
  {volume} {72}},\ \bibinfo {pages} {113003} (\bibinfo {year} {2005})},\
  \Eprint {https://arxiv.org/abs/hep-ph/0511315} {arXiv:hep-ph/0511315}
  \BibitemShut {NoStop}%
\bibitem [{\citenamefont {Flores-Baez}\ \emph {et~al.}(2006)\citenamefont
  {Flores-Baez}, \citenamefont {Flores-Tlalpa}, \citenamefont {Lopez~Castro},\
  and\ \citenamefont {Toledo~Sanchez}}]{Flores-Baez:2006yiq}%
  \BibitemOpen
  \bibfield  {author} {\bibinfo {author} {\bibfnamefont {F.}~\bibnamefont
  {Flores-Baez}}, \bibinfo {author} {\bibfnamefont {A.}~\bibnamefont
  {Flores-Tlalpa}}, \bibinfo {author} {\bibfnamefont {G.}~\bibnamefont
  {Lopez~Castro}},\ and\ \bibinfo {author} {\bibfnamefont {G.}~\bibnamefont
  {Toledo~Sanchez}},\ }\href {https://doi.org/10.1103/PhysRevD.74.071301}
  {\bibfield  {journal} {\bibinfo  {journal} {Phys. Rev. D}\ }\textbf {\bibinfo
  {volume} {74}},\ \bibinfo {pages} {071301} (\bibinfo {year} {2006})},\
  \Eprint {https://arxiv.org/abs/hep-ph/0608084} {arXiv:hep-ph/0608084}
  \BibitemShut {NoStop}%
\bibitem [{\citenamefont {Miranda}\ and\ \citenamefont
  {Roig}(2020)}]{Miranda:2020wdg}%
  \BibitemOpen
  \bibfield  {author} {\bibinfo {author} {\bibfnamefont {J.~A.}\ \bibnamefont
  {Miranda}}\ and\ \bibinfo {author} {\bibfnamefont {P.}~\bibnamefont {Roig}},\
  }\href {https://doi.org/10.1103/PhysRevD.102.114017} {\bibfield  {journal}
  {\bibinfo  {journal} {Phys. Rev. D}\ }\textbf {\bibinfo {volume} {102}},\
  \bibinfo {pages} {114017} (\bibinfo {year} {2020})},\ \Eprint
  {https://arxiv.org/abs/2007.11019} {arXiv:2007.11019 [hep-ph]} \BibitemShut
  {NoStop}%
\bibitem [{\citenamefont {Castro}\ \emph {et~al.}(2025)\citenamefont {Castro},
  \citenamefont {Miranda},\ and\ \citenamefont {Roig}}]{Castro:2024prg}%
  \BibitemOpen
  \bibfield  {author} {\bibinfo {author} {\bibfnamefont {G.~L.}\ \bibnamefont
  {Castro}}, \bibinfo {author} {\bibfnamefont {A.}~\bibnamefont {Miranda}},\
  and\ \bibinfo {author} {\bibfnamefont {P.}~\bibnamefont {Roig}},\ }\href
  {https://doi.org/10.1103/PhysRevD.111.073004} {\bibfield  {journal} {\bibinfo
   {journal} {Phys. Rev. D}\ }\textbf {\bibinfo {volume} {111}},\ \bibinfo
  {pages} {073004} (\bibinfo {year} {2025})},\ \Eprint
  {https://arxiv.org/abs/2411.07696} {arXiv:2411.07696 [hep-ph]} \BibitemShut
  {NoStop}%
\bibitem [{\citenamefont {Kampf}\ and\ \citenamefont
  {Novotny}(2011)}]{Kampf:2011ty}%
  \BibitemOpen
  \bibfield  {author} {\bibinfo {author} {\bibfnamefont {K.}~\bibnamefont
  {Kampf}}\ and\ \bibinfo {author} {\bibfnamefont {J.}~\bibnamefont
  {Novotny}},\ }\href {https://doi.org/10.1103/PhysRevD.84.014036} {\bibfield
  {journal} {\bibinfo  {journal} {Phys. Rev. D}\ }\textbf {\bibinfo {volume}
  {84}},\ \bibinfo {pages} {014036} (\bibinfo {year} {2011})},\ \Eprint
  {https://arxiv.org/abs/1104.3137} {arXiv:1104.3137 [hep-ph]} \BibitemShut
  {NoStop}%
\bibitem [{\citenamefont {Chen}\ \emph {et~al.}(2022)\citenamefont {Chen},
  \citenamefont {Duan},\ and\ \citenamefont {Guo}}]{Chen:2022nxm}%
  \BibitemOpen
  \bibfield  {author} {\bibinfo {author} {\bibfnamefont {C.}~\bibnamefont
  {Chen}}, \bibinfo {author} {\bibfnamefont {C.-G.}\ \bibnamefont {Duan}},\
  and\ \bibinfo {author} {\bibfnamefont {Z.-H.}\ \bibnamefont {Guo}},\ }\href
  {https://doi.org/10.1007/JHEP08(2022)144} {\bibfield  {journal} {\bibinfo
  {journal} {JHEP}\ }\textbf {\bibinfo {volume} {08}},\ \bibinfo {pages}
  {144}},\ \Eprint {https://arxiv.org/abs/2201.12764} {arXiv:2201.12764
  [hep-ph]} \BibitemShut {NoStop}%
\bibitem [{\citenamefont {Ruiz-Femenia}\ \emph {et~al.}(2003)\citenamefont
  {Ruiz-Femenia}, \citenamefont {Pich},\ and\ \citenamefont
  {Portoles}}]{Ruiz-Femenia:2003jdx}%
  \BibitemOpen
  \bibfield  {author} {\bibinfo {author} {\bibfnamefont {P.~D.}\ \bibnamefont
  {Ruiz-Femenia}}, \bibinfo {author} {\bibfnamefont {A.}~\bibnamefont {Pich}},\
  and\ \bibinfo {author} {\bibfnamefont {J.}~\bibnamefont {Portoles}},\ }\href
  {https://doi.org/10.1088/1126-6708/2003/07/003} {\bibfield  {journal}
  {\bibinfo  {journal} {JHEP}\ }\textbf {\bibinfo {volume} {07}},\ \bibinfo
  {pages} {003}},\ \Eprint {https://arxiv.org/abs/hep-ph/0306157}
  {arXiv:hep-ph/0306157} \BibitemShut {NoStop}%
\bibitem [{\citenamefont {Fujikawa}\ \emph {et~al.}(2008)\citenamefont
  {Fujikawa} \emph {et~al.}}]{Belle:2008xpe}%
  \BibitemOpen
  \bibfield  {author} {\bibinfo {author} {\bibfnamefont {M.}~\bibnamefont
  {Fujikawa}} \emph {et~al.} (\bibinfo {collaboration} {Belle}),\ }\href
  {https://doi.org/10.1103/PhysRevD.78.072006} {\bibfield  {journal} {\bibinfo
  {journal} {Phys. Rev. D}\ }\textbf {\bibinfo {volume} {78}},\ \bibinfo
  {pages} {072006} (\bibinfo {year} {2008})},\ \Eprint
  {https://arxiv.org/abs/0805.3773} {arXiv:0805.3773 [hep-ex]} \BibitemShut
  {NoStop}%
\bibitem [{\citenamefont {Davier}\ \emph {et~al.}(2014)\citenamefont {Davier},
  \citenamefont {H\"ocker}, \citenamefont {Malaescu}, \citenamefont {Yuan},\
  and\ \citenamefont {Zhang}}]{Davier:2013sfa}%
  \BibitemOpen
  \bibfield  {author} {\bibinfo {author} {\bibfnamefont {M.}~\bibnamefont
  {Davier}}, \bibinfo {author} {\bibfnamefont {A.}~\bibnamefont {H\"ocker}},
  \bibinfo {author} {\bibfnamefont {B.}~\bibnamefont {Malaescu}}, \bibinfo
  {author} {\bibfnamefont {C.-Z.}\ \bibnamefont {Yuan}},\ and\ \bibinfo
  {author} {\bibfnamefont {Z.}~\bibnamefont {Zhang}},\ }\href
  {https://doi.org/10.1140/epjc/s10052-014-2803-9} {\bibfield  {journal}
  {\bibinfo  {journal} {Eur. Phys. J. C}\ }\textbf {\bibinfo {volume} {74}},\
  \bibinfo {pages} {2803} (\bibinfo {year} {2014})},\ \Eprint
  {https://arxiv.org/abs/1312.1501} {arXiv:1312.1501 [hep-ex]} \BibitemShut
  {NoStop}%
\bibitem [{\citenamefont {Anderson}\ \emph {et~al.}(2000)\citenamefont
  {Anderson} \emph {et~al.}}]{CLEO:1999dln}%
  \BibitemOpen
  \bibfield  {author} {\bibinfo {author} {\bibfnamefont {S.}~\bibnamefont
  {Anderson}} \emph {et~al.} (\bibinfo {collaboration} {CLEO}),\ }\href
  {https://doi.org/10.1103/PhysRevD.61.112002} {\bibfield  {journal} {\bibinfo
  {journal} {Phys. Rev. D}\ }\textbf {\bibinfo {volume} {61}},\ \bibinfo
  {pages} {112002} (\bibinfo {year} {2000})},\ \Eprint
  {https://arxiv.org/abs/hep-ex/9910046} {arXiv:hep-ex/9910046} \BibitemShut
  {NoStop}%
\bibitem [{\citenamefont {Ackerstaff}\ \emph {et~al.}(1999)\citenamefont
  {Ackerstaff} \emph {et~al.}}]{OPAL:1998rrm}%
  \BibitemOpen
  \bibfield  {author} {\bibinfo {author} {\bibfnamefont {K.}~\bibnamefont
  {Ackerstaff}} \emph {et~al.} (\bibinfo {collaboration} {OPAL}),\ }\href
  {https://doi.org/10.1007/s100529901061} {\bibfield  {journal} {\bibinfo
  {journal} {Eur. Phys. J. C}\ }\textbf {\bibinfo {volume} {7}},\ \bibinfo
  {pages} {571} (\bibinfo {year} {1999})},\ \Eprint
  {https://arxiv.org/abs/hep-ex/9808019} {arXiv:hep-ex/9808019} \BibitemShut
  {NoStop}%
\bibitem [{\citenamefont {Wess}\ and\ \citenamefont
  {Zumino}(1971)}]{Wess:1971yu}%
  \BibitemOpen
  \bibfield  {author} {\bibinfo {author} {\bibfnamefont {J.}~\bibnamefont
  {Wess}}\ and\ \bibinfo {author} {\bibfnamefont {B.}~\bibnamefont {Zumino}},\
  }\href {https://doi.org/10.1016/0370-2693(71)90582-X} {\bibfield  {journal}
  {\bibinfo  {journal} {Phys. Lett. B}\ }\textbf {\bibinfo {volume} {37}},\
  \bibinfo {pages} {95} (\bibinfo {year} {1971})}\BibitemShut {NoStop}%
\bibitem [{\citenamefont {Witten}(1983)}]{Witten:1983tw}%
  \BibitemOpen
  \bibfield  {author} {\bibinfo {author} {\bibfnamefont {E.}~\bibnamefont
  {Witten}},\ }\href {https://doi.org/10.1016/0550-3213(83)90063-9} {\bibfield
  {journal} {\bibinfo  {journal} {Nucl. Phys. B}\ }\textbf {\bibinfo {volume}
  {223}},\ \bibinfo {pages} {422} (\bibinfo {year} {1983})}\BibitemShut
  {NoStop}%
\bibitem [{\citenamefont {Cirigliano}\ \emph {et~al.}(2006)\citenamefont
  {Cirigliano}, \citenamefont {Ecker}, \citenamefont {Eidemuller},
  \citenamefont {Kaiser}, \citenamefont {Pich},\ and\ \citenamefont
  {Portoles}}]{Cirigliano:2006hb}%
  \BibitemOpen
  \bibfield  {author} {\bibinfo {author} {\bibfnamefont {V.}~\bibnamefont
  {Cirigliano}}, \bibinfo {author} {\bibfnamefont {G.}~\bibnamefont {Ecker}},
  \bibinfo {author} {\bibfnamefont {M.}~\bibnamefont {Eidemuller}}, \bibinfo
  {author} {\bibfnamefont {R.}~\bibnamefont {Kaiser}}, \bibinfo {author}
  {\bibfnamefont {A.}~\bibnamefont {Pich}},\ and\ \bibinfo {author}
  {\bibfnamefont {J.}~\bibnamefont {Portoles}},\ }\href
  {https://doi.org/10.1016/j.nuclphysb.2006.07.010} {\bibfield  {journal}
  {\bibinfo  {journal} {Nucl. Phys. B}\ }\textbf {\bibinfo {volume} {753}},\
  \bibinfo {pages} {139} (\bibinfo {year} {2006})},\ \Eprint
  {https://arxiv.org/abs/hep-ph/0603205} {arXiv:hep-ph/0603205} \BibitemShut
  {NoStop}%
\bibitem [{\citenamefont {Chen}\ \emph {et~al.}(2012)\citenamefont {Chen},
  \citenamefont {Guo},\ and\ \citenamefont {Zheng}}]{Chen:2012vw}%
  \BibitemOpen
  \bibfield  {author} {\bibinfo {author} {\bibfnamefont {Y.-H.}\ \bibnamefont
  {Chen}}, \bibinfo {author} {\bibfnamefont {Z.-H.}\ \bibnamefont {Guo}},\ and\
  \bibinfo {author} {\bibfnamefont {H.-Q.}\ \bibnamefont {Zheng}},\ }\href
  {https://doi.org/10.1103/PhysRevD.85.054018} {\bibfield  {journal} {\bibinfo
  {journal} {Phys. Rev. D}\ }\textbf {\bibinfo {volume} {85}},\ \bibinfo
  {pages} {054018} (\bibinfo {year} {2012})},\ \Eprint
  {https://arxiv.org/abs/1201.2135} {arXiv:1201.2135 [hep-ph]} \BibitemShut
  {NoStop}%
\bibitem [{\citenamefont {Chen}\ \emph {et~al.}(2014)\citenamefont {Chen},
  \citenamefont {Guo},\ and\ \citenamefont {Zheng}}]{Chen:2013nna}%
  \BibitemOpen
  \bibfield  {author} {\bibinfo {author} {\bibfnamefont {Y.-H.}\ \bibnamefont
  {Chen}}, \bibinfo {author} {\bibfnamefont {Z.-H.}\ \bibnamefont {Guo}},\ and\
  \bibinfo {author} {\bibfnamefont {H.-Q.}\ \bibnamefont {Zheng}},\ }\href
  {https://doi.org/10.1103/PhysRevD.90.034013} {\bibfield  {journal} {\bibinfo
  {journal} {Phys. Rev. D}\ }\textbf {\bibinfo {volume} {90}},\ \bibinfo
  {pages} {034013} (\bibinfo {year} {2014})},\ \Eprint
  {https://arxiv.org/abs/1311.3366} {arXiv:1311.3366 [hep-ph]} \BibitemShut
  {NoStop}%
\bibitem [{\citenamefont {Chen}\ \emph {et~al.}(2015)\citenamefont {Chen},
  \citenamefont {Guo},\ and\ \citenamefont {Zou}}]{Chen:2014yta}%
  \BibitemOpen
  \bibfield  {author} {\bibinfo {author} {\bibfnamefont {Y.-H.}\ \bibnamefont
  {Chen}}, \bibinfo {author} {\bibfnamefont {Z.-H.}\ \bibnamefont {Guo}},\ and\
  \bibinfo {author} {\bibfnamefont {B.-S.}\ \bibnamefont {Zou}},\ }\href
  {https://doi.org/10.1103/PhysRevD.91.014010} {\bibfield  {journal} {\bibinfo
  {journal} {Phys. Rev. D}\ }\textbf {\bibinfo {volume} {91}},\ \bibinfo
  {pages} {014010} (\bibinfo {year} {2015})},\ \Eprint
  {https://arxiv.org/abs/1411.1159} {arXiv:1411.1159 [hep-ph]} \BibitemShut
  {NoStop}%
\bibitem [{\citenamefont {Yan}\ \emph {et~al.}(2023)\citenamefont {Yan},
  \citenamefont {Chen}, \citenamefont {Duan},\ and\ \citenamefont
  {Guo}}]{Yan:2023nqz}%
  \BibitemOpen
  \bibfield  {author} {\bibinfo {author} {\bibfnamefont {L.-W.}\ \bibnamefont
  {Yan}}, \bibinfo {author} {\bibfnamefont {Y.-H.}\ \bibnamefont {Chen}},
  \bibinfo {author} {\bibfnamefont {C.-G.}\ \bibnamefont {Duan}},\ and\
  \bibinfo {author} {\bibfnamefont {Z.-H.}\ \bibnamefont {Guo}},\ }\href
  {https://doi.org/10.1103/PhysRevD.107.034022} {\bibfield  {journal} {\bibinfo
   {journal} {Phys. Rev. D}\ }\textbf {\bibinfo {volume} {107}},\ \bibinfo
  {pages} {034022} (\bibinfo {year} {2023})},\ \Eprint
  {https://arxiv.org/abs/2301.03869} {arXiv:2301.03869 [hep-ph]} \BibitemShut
  {NoStop}%
\bibitem [{\citenamefont {Guo}(2008)}]{Guo:2008sh}%
  \BibitemOpen
  \bibfield  {author} {\bibinfo {author} {\bibfnamefont {Z.-H.}\ \bibnamefont
  {Guo}},\ }\href {https://doi.org/10.1103/PhysRevD.78.033004} {\bibfield
  {journal} {\bibinfo  {journal} {Phys. Rev. D}\ }\textbf {\bibinfo {volume}
  {78}},\ \bibinfo {pages} {033004} (\bibinfo {year} {2008})},\ \Eprint
  {https://arxiv.org/abs/0806.4322} {arXiv:0806.4322 [hep-ph]} \BibitemShut
  {NoStop}%
\bibitem [{\citenamefont {Guo}\ and\ \citenamefont {Roig}(2010)}]{Guo:2010dv}%
  \BibitemOpen
  \bibfield  {author} {\bibinfo {author} {\bibfnamefont {Z.-H.}\ \bibnamefont
  {Guo}}\ and\ \bibinfo {author} {\bibfnamefont {P.}~\bibnamefont {Roig}},\
  }\href {https://doi.org/10.1103/PhysRevD.82.113016} {\bibfield  {journal}
  {\bibinfo  {journal} {Phys. Rev. D}\ }\textbf {\bibinfo {volume} {82}},\
  \bibinfo {pages} {113016} (\bibinfo {year} {2010})},\ \Eprint
  {https://arxiv.org/abs/1009.2542} {arXiv:1009.2542 [hep-ph]} \BibitemShut
  {NoStop}%
\bibitem [{\citenamefont {Ecker}\ \emph
  {et~al.}(1989{\natexlab{b}})\citenamefont {Ecker}, \citenamefont {Gasser},
  \citenamefont {Leutwyler}, \citenamefont {Pich},\ and\ \citenamefont
  {de~Rafael}}]{Ecker:1989yg}%
  \BibitemOpen
  \bibfield  {author} {\bibinfo {author} {\bibfnamefont {G.}~\bibnamefont
  {Ecker}}, \bibinfo {author} {\bibfnamefont {J.}~\bibnamefont {Gasser}},
  \bibinfo {author} {\bibfnamefont {H.}~\bibnamefont {Leutwyler}}, \bibinfo
  {author} {\bibfnamefont {A.}~\bibnamefont {Pich}},\ and\ \bibinfo {author}
  {\bibfnamefont {E.}~\bibnamefont {de~Rafael}},\ }\href
  {https://doi.org/10.1016/0370-2693(89)91627-4} {\bibfield  {journal}
  {\bibinfo  {journal} {Phys. Lett. B}\ }\textbf {\bibinfo {volume} {223}},\
  \bibinfo {pages} {425} (\bibinfo {year} {1989}{\natexlab{b}})}\BibitemShut
  {NoStop}%
\bibitem [{\citenamefont {Roig}\ and\ \citenamefont
  {Sanz~Cillero}(2014)}]{Roig:2013baa}%
  \BibitemOpen
  \bibfield  {author} {\bibinfo {author} {\bibfnamefont {P.}~\bibnamefont
  {Roig}}\ and\ \bibinfo {author} {\bibfnamefont {J.~J.}\ \bibnamefont
  {Sanz~Cillero}},\ }\href {https://doi.org/10.1016/j.physletb.2014.04.034}
  {\bibfield  {journal} {\bibinfo  {journal} {Phys. Lett. B}\ }\textbf
  {\bibinfo {volume} {733}},\ \bibinfo {pages} {158} (\bibinfo {year}
  {2014})},\ \Eprint {https://arxiv.org/abs/1312.6206} {arXiv:1312.6206
  [hep-ph]} \BibitemShut {NoStop}%
\bibitem [{\citenamefont {Bramon}\ \emph {et~al.}(2001)\citenamefont {Bramon},
  \citenamefont {Escribano}, \citenamefont {Lucio~Martinez},\ and\
  \citenamefont {Napsuciale}}]{Bramon:2001un}%
  \BibitemOpen
  \bibfield  {author} {\bibinfo {author} {\bibfnamefont {A.}~\bibnamefont
  {Bramon}}, \bibinfo {author} {\bibfnamefont {R.}~\bibnamefont {Escribano}},
  \bibinfo {author} {\bibfnamefont {J.~L.}\ \bibnamefont {Lucio~Martinez}},\
  and\ \bibinfo {author} {\bibfnamefont {M.}~\bibnamefont {Napsuciale}},\
  }\href {https://doi.org/10.1016/S0370-2693(01)01007-3} {\bibfield  {journal}
  {\bibinfo  {journal} {Phys. Lett. B}\ }\textbf {\bibinfo {volume} {517}},\
  \bibinfo {pages} {345} (\bibinfo {year} {2001})},\ \Eprint
  {https://arxiv.org/abs/hep-ph/0105179} {arXiv:hep-ph/0105179} \BibitemShut
  {NoStop}%
\bibitem [{\citenamefont {Escribano}(2006)}]{Escribano:2006mb}%
  \BibitemOpen
  \bibfield  {author} {\bibinfo {author} {\bibfnamefont {R.}~\bibnamefont
  {Escribano}},\ }\href {https://doi.org/10.1103/PhysRevD.74.114020} {\bibfield
   {journal} {\bibinfo  {journal} {Phys. Rev. D}\ }\textbf {\bibinfo {volume}
  {74}},\ \bibinfo {pages} {114020} (\bibinfo {year} {2006})},\ \Eprint
  {https://arxiv.org/abs/hep-ph/0606314} {arXiv:hep-ph/0606314} \BibitemShut
  {NoStop}%
\bibitem [{\citenamefont {Oh}\ and\ \citenamefont {Kim}(2003)}]{Oh:2003zz}%
  \BibitemOpen
  \bibfield  {author} {\bibinfo {author} {\bibfnamefont {Y.-s.}\ \bibnamefont
  {Oh}}\ and\ \bibinfo {author} {\bibfnamefont {H.-c.}\ \bibnamefont {Kim}},\
  }\href {https://doi.org/10.1103/PhysRevD.68.094003} {\bibfield  {journal}
  {\bibinfo  {journal} {Phys. Rev. D}\ }\textbf {\bibinfo {volume} {68}},\
  \bibinfo {pages} {094003} (\bibinfo {year} {2003})},\ \Eprint
  {https://arxiv.org/abs/hep-ph/0307286} {arXiv:hep-ph/0307286} \BibitemShut
  {NoStop}%
\bibitem [{\citenamefont {Gokalp}\ and\ \citenamefont
  {Yilmaz}(2000)}]{Gokalp:2000xy}%
  \BibitemOpen
  \bibfield  {author} {\bibinfo {author} {\bibfnamefont {A.}~\bibnamefont
  {Gokalp}}\ and\ \bibinfo {author} {\bibfnamefont {O.}~\bibnamefont
  {Yilmaz}},\ }\href {https://doi.org/10.1016/S0370-2693(00)01126-6} {\bibfield
   {journal} {\bibinfo  {journal} {Phys. Lett. B}\ }\textbf {\bibinfo {volume}
  {494}},\ \bibinfo {pages} {69} (\bibinfo {year} {2000})},\ \Eprint
  {https://arxiv.org/abs/nucl-th/0008011} {arXiv:nucl-th/0008011} \BibitemShut
  {NoStop}%
\bibitem [{\citenamefont {Eidelman}\ \emph {et~al.}(2010)\citenamefont
  {Eidelman}, \citenamefont {Ivashyn}, \citenamefont {Korchin}, \citenamefont
  {Pancheri},\ and\ \citenamefont {Shekhovtsova}}]{Eidelman:2010ta}%
  \BibitemOpen
  \bibfield  {author} {\bibinfo {author} {\bibfnamefont {S.}~\bibnamefont
  {Eidelman}}, \bibinfo {author} {\bibfnamefont {S.}~\bibnamefont {Ivashyn}},
  \bibinfo {author} {\bibfnamefont {A.}~\bibnamefont {Korchin}}, \bibinfo
  {author} {\bibfnamefont {G.}~\bibnamefont {Pancheri}},\ and\ \bibinfo
  {author} {\bibfnamefont {O.}~\bibnamefont {Shekhovtsova}},\ }\href
  {https://doi.org/10.1140/epjc/s10052-010-1394-3} {\bibfield  {journal}
  {\bibinfo  {journal} {Eur. Phys. J. C}\ }\textbf {\bibinfo {volume} {69}},\
  \bibinfo {pages} {103} (\bibinfo {year} {2010})},\ \Eprint
  {https://arxiv.org/abs/1003.2141} {arXiv:1003.2141 [hep-ph]} \BibitemShut
  {NoStop}%
\bibitem [{\citenamefont {Oller}(1998)}]{Oller:1998ia}%
  \BibitemOpen
  \bibfield  {author} {\bibinfo {author} {\bibfnamefont {J.~A.}\ \bibnamefont
  {Oller}},\ }\href {https://doi.org/10.1016/S0370-2693(98)00298-6} {\bibfield
  {journal} {\bibinfo  {journal} {Phys. Lett. B}\ }\textbf {\bibinfo {volume}
  {426}},\ \bibinfo {pages} {7} (\bibinfo {year} {1998})},\ \Eprint
  {https://arxiv.org/abs/hep-ph/9803214} {arXiv:hep-ph/9803214} \BibitemShut
  {NoStop}%
\bibitem [{\citenamefont {Oller}(2003)}]{Oller:2002na}%
  \BibitemOpen
  \bibfield  {author} {\bibinfo {author} {\bibfnamefont {J.~A.}\ \bibnamefont
  {Oller}},\ }\href {https://doi.org/10.1016/S0375-9474(02)01360-X} {\bibfield
  {journal} {\bibinfo  {journal} {Nucl. Phys. A}\ }\textbf {\bibinfo {volume}
  {714}},\ \bibinfo {pages} {161} (\bibinfo {year} {2003})},\ \Eprint
  {https://arxiv.org/abs/hep-ph/0205121} {arXiv:hep-ph/0205121} \BibitemShut
  {NoStop}%
\bibitem [{\citenamefont {Palomar}\ \emph {et~al.}(2002)\citenamefont
  {Palomar}, \citenamefont {Hirenzaki},\ and\ \citenamefont
  {Oset}}]{Palomar:2001vg}%
  \BibitemOpen
  \bibfield  {author} {\bibinfo {author} {\bibfnamefont {J.~E.}\ \bibnamefont
  {Palomar}}, \bibinfo {author} {\bibfnamefont {S.}~\bibnamefont {Hirenzaki}},\
  and\ \bibinfo {author} {\bibfnamefont {E.}~\bibnamefont {Oset}},\ }\href
  {https://doi.org/10.1016/S0375-9474(02)00960-0} {\bibfield  {journal}
  {\bibinfo  {journal} {Nucl. Phys. A}\ }\textbf {\bibinfo {volume} {707}},\
  \bibinfo {pages} {161} (\bibinfo {year} {2002})},\ \Eprint
  {https://arxiv.org/abs/hep-ph/0111308} {arXiv:hep-ph/0111308} \BibitemShut
  {NoStop}%
\bibitem [{\citenamefont {Close}\ \emph {et~al.}(1993)\citenamefont {Close},
  \citenamefont {Isgur},\ and\ \citenamefont {Kumano}}]{Close:1992ay}%
  \BibitemOpen
  \bibfield  {author} {\bibinfo {author} {\bibfnamefont {F.~E.}\ \bibnamefont
  {Close}}, \bibinfo {author} {\bibfnamefont {N.}~\bibnamefont {Isgur}},\ and\
  \bibinfo {author} {\bibfnamefont {S.}~\bibnamefont {Kumano}},\ }\href
  {https://doi.org/10.1016/0550-3213(93)90329-N} {\bibfield  {journal}
  {\bibinfo  {journal} {Nucl. Phys. B}\ }\textbf {\bibinfo {volume} {389}},\
  \bibinfo {pages} {513} (\bibinfo {year} {1993})},\ \Eprint
  {https://arxiv.org/abs/hep-ph/9301253} {arXiv:hep-ph/9301253} \BibitemShut
  {NoStop}%
\bibitem [{\citenamefont {Oller}\ and\ \citenamefont
  {Oset}(1997)}]{Oller:1997ti}%
  \BibitemOpen
  \bibfield  {author} {\bibinfo {author} {\bibfnamefont {J.~A.}\ \bibnamefont
  {Oller}}\ and\ \bibinfo {author} {\bibfnamefont {E.}~\bibnamefont {Oset}},\
  }\href {https://doi.org/10.1016/S0375-9474(97)00160-7} {\bibfield  {journal}
  {\bibinfo  {journal} {Nucl. Phys. A}\ }\textbf {\bibinfo {volume} {620}},\
  \bibinfo {pages} {438} (\bibinfo {year} {1997})},\ \bibinfo {note} {[Erratum:
  Nucl.Phys.A 652, 407--409 (1999)]},\ \Eprint
  {https://arxiv.org/abs/hep-ph/9702314} {arXiv:hep-ph/9702314} \BibitemShut
  {NoStop}%
\bibitem [{\citenamefont {Oller}\ \emph {et~al.}(1999)\citenamefont {Oller},
  \citenamefont {Oset},\ and\ \citenamefont {Pelaez}}]{Oller:1998hw}%
  \BibitemOpen
  \bibfield  {author} {\bibinfo {author} {\bibfnamefont {J.~A.}\ \bibnamefont
  {Oller}}, \bibinfo {author} {\bibfnamefont {E.}~\bibnamefont {Oset}},\ and\
  \bibinfo {author} {\bibfnamefont {J.~R.}\ \bibnamefont {Pelaez}},\ }\href
  {https://doi.org/10.1103/PhysRevD.59.074001} {\bibfield  {journal} {\bibinfo
  {journal} {Phys. Rev. D}\ }\textbf {\bibinfo {volume} {59}},\ \bibinfo
  {pages} {074001} (\bibinfo {year} {1999})},\ \bibinfo {note} {[Erratum:
  Phys.Rev.D 60, 099906 (1999), Erratum: Phys.Rev.D 75, 099903 (2007)]},\
  \Eprint {https://arxiv.org/abs/hep-ph/9804209} {arXiv:hep-ph/9804209}
  \BibitemShut {NoStop}%
\bibitem [{\citenamefont {Oller}\ and\ \citenamefont
  {Oset}(1999)}]{Oller:1998zr}%
  \BibitemOpen
  \bibfield  {author} {\bibinfo {author} {\bibfnamefont {J.~A.}\ \bibnamefont
  {Oller}}\ and\ \bibinfo {author} {\bibfnamefont {E.}~\bibnamefont {Oset}},\
  }\href {https://doi.org/10.1103/PhysRevD.60.074023} {\bibfield  {journal}
  {\bibinfo  {journal} {Phys. Rev. D}\ }\textbf {\bibinfo {volume} {60}},\
  \bibinfo {pages} {074023} (\bibinfo {year} {1999})},\ \Eprint
  {https://arxiv.org/abs/hep-ph/9809337} {arXiv:hep-ph/9809337} \BibitemShut
  {NoStop}%
\bibitem [{\citenamefont {Guo}\ and\ \citenamefont {Oller}(2011)}]{Guo:2011pa}%
  \BibitemOpen
  \bibfield  {author} {\bibinfo {author} {\bibfnamefont {Z.-H.}\ \bibnamefont
  {Guo}}\ and\ \bibinfo {author} {\bibfnamefont {J.~A.}\ \bibnamefont
  {Oller}},\ }\href {https://doi.org/10.1103/PhysRevD.84.034005} {\bibfield
  {journal} {\bibinfo  {journal} {Phys. Rev. D}\ }\textbf {\bibinfo {volume}
  {84}},\ \bibinfo {pages} {034005} (\bibinfo {year} {2011})},\ \Eprint
  {https://arxiv.org/abs/1104.2849} {arXiv:1104.2849 [hep-ph]} \BibitemShut
  {NoStop}%
\bibitem [{\citenamefont {Guo}\ \emph {et~al.}(2012)\citenamefont {Guo},
  \citenamefont {Oller},\ and\ \citenamefont {Ruiz~de Elvira}}]{Guo:2012yt}%
  \BibitemOpen
  \bibfield  {author} {\bibinfo {author} {\bibfnamefont {Z.-H.}\ \bibnamefont
  {Guo}}, \bibinfo {author} {\bibfnamefont {J.~A.}\ \bibnamefont {Oller}},\
  and\ \bibinfo {author} {\bibfnamefont {J.}~\bibnamefont {Ruiz~de Elvira}},\
  }\href {https://doi.org/10.1103/PhysRevD.86.054006} {\bibfield  {journal}
  {\bibinfo  {journal} {Phys. Rev. D}\ }\textbf {\bibinfo {volume} {86}},\
  \bibinfo {pages} {054006} (\bibinfo {year} {2012})},\ \Eprint
  {https://arxiv.org/abs/1206.4163} {arXiv:1206.4163 [hep-ph]} \BibitemShut
  {NoStop}%
\bibitem [{\citenamefont {Gao}\ \emph {et~al.}(2019)\citenamefont {Gao},
  \citenamefont {Guo},\ and\ \citenamefont {Pang}}]{Gao:2019idb}%
  \BibitemOpen
  \bibfield  {author} {\bibinfo {author} {\bibfnamefont {R.}~\bibnamefont
  {Gao}}, \bibinfo {author} {\bibfnamefont {Z.-H.}\ \bibnamefont {Guo}},\ and\
  \bibinfo {author} {\bibfnamefont {J.-Y.}\ \bibnamefont {Pang}},\ }\href
  {https://doi.org/10.1103/PhysRevD.100.114028} {\bibfield  {journal} {\bibinfo
   {journal} {Phys. Rev. D}\ }\textbf {\bibinfo {volume} {100}},\ \bibinfo
  {pages} {114028} (\bibinfo {year} {2019})},\ \Eprint
  {https://arxiv.org/abs/1907.01787} {arXiv:1907.01787 [hep-ph]} \BibitemShut
  {NoStop}%
\bibitem [{\citenamefont {Gasser}\ and\ \citenamefont
  {Leutwyler}(1985)}]{Gasser:1984gg}%
  \BibitemOpen
  \bibfield  {author} {\bibinfo {author} {\bibfnamefont {J.}~\bibnamefont
  {Gasser}}\ and\ \bibinfo {author} {\bibfnamefont {H.}~\bibnamefont
  {Leutwyler}},\ }\href {https://doi.org/10.1016/0550-3213(85)90492-4}
  {\bibfield  {journal} {\bibinfo  {journal} {Nucl. Phys. B}\ }\textbf
  {\bibinfo {volume} {250}},\ \bibinfo {pages} {465} (\bibinfo {year}
  {1985})}\BibitemShut {NoStop}%
\bibitem [{\citenamefont {Pelaez}(2016)}]{Pelaez:2015qba}%
  \BibitemOpen
  \bibfield  {author} {\bibinfo {author} {\bibfnamefont {J.~R.}\ \bibnamefont
  {Pelaez}},\ }\href {https://doi.org/10.1016/j.physrep.2016.09.001} {\bibfield
   {journal} {\bibinfo  {journal} {Phys. Rept.}\ }\textbf {\bibinfo {volume}
  {658}},\ \bibinfo {pages} {1} (\bibinfo {year} {2016})},\ \Eprint
  {https://arxiv.org/abs/1510.00653} {arXiv:1510.00653 [hep-ph]} \BibitemShut
  {NoStop}%
\bibitem [{\citenamefont {Navas}\ \emph {et~al.}(2024)\citenamefont {Navas}
  \emph {et~al.}}]{ParticleDataGroup:2024cfk}%
  \BibitemOpen
  \bibfield  {author} {\bibinfo {author} {\bibfnamefont {S.}~\bibnamefont
  {Navas}} \emph {et~al.} (\bibinfo {collaboration} {Particle Data Group}),\
  }\href {https://doi.org/10.1103/PhysRevD.110.030001} {\bibfield  {journal}
  {\bibinfo  {journal} {Phys. Rev. D}\ }\textbf {\bibinfo {volume} {110}},\
  \bibinfo {pages} {030001} (\bibinfo {year} {2024})}\BibitemShut {NoStop}%
\bibitem [{\citenamefont {Low}(1958)}]{Low:1958sn}%
  \BibitemOpen
  \bibfield  {author} {\bibinfo {author} {\bibfnamefont {F.~E.}\ \bibnamefont
  {Low}},\ }\href {https://doi.org/10.1103/PhysRev.110.974} {\bibfield
  {journal} {\bibinfo  {journal} {Phys. Rev.}\ }\textbf {\bibinfo {volume}
  {110}},\ \bibinfo {pages} {974} (\bibinfo {year} {1958})}\BibitemShut
  {NoStop}%
\bibitem [{\citenamefont {Byckling}\ and\ \citenamefont
  {Kajantie}(1971)}]{Byckling:1971vca}%
  \BibitemOpen
  \bibfield  {author} {\bibinfo {author} {\bibfnamefont {E.}~\bibnamefont
  {Byckling}}\ and\ \bibinfo {author} {\bibfnamefont {K.}~\bibnamefont
  {Kajantie}},\ }\href@noop {} {\emph {\bibinfo {title} {{Particle Kinematics}:
  {(Chapters I-VI, X)}}}}\ (\bibinfo  {publisher} {University of Jyvaskyla},\
  \bibinfo {address} {Jyvaskyla, Finland},\ \bibinfo {year} {1971})\BibitemShut
  {NoStop}%
\bibitem [{\citenamefont {Gourdin}\ and\ \citenamefont
  {De~Rafael}(1969)}]{Gourdin:1969dm}%
  \BibitemOpen
  \bibfield  {author} {\bibinfo {author} {\bibfnamefont {M.}~\bibnamefont
  {Gourdin}}\ and\ \bibinfo {author} {\bibfnamefont {E.}~\bibnamefont
  {De~Rafael}},\ }\href {https://doi.org/10.1016/0550-3213(69)90333-2}
  {\bibfield  {journal} {\bibinfo  {journal} {Nucl. Phys. B}\ }\textbf
  {\bibinfo {volume} {10}},\ \bibinfo {pages} {667} (\bibinfo {year}
  {1969})}\BibitemShut {NoStop}%
\bibitem [{\citenamefont {Brodsky}\ and\ \citenamefont
  {De~Rafael}(1968)}]{Brodsky:1967sr}%
  \BibitemOpen
  \bibfield  {author} {\bibinfo {author} {\bibfnamefont {S.~J.}\ \bibnamefont
  {Brodsky}}\ and\ \bibinfo {author} {\bibfnamefont {E.}~\bibnamefont
  {De~Rafael}},\ }\href {https://doi.org/10.1103/PhysRev.168.1620} {\bibfield
  {journal} {\bibinfo  {journal} {Phys. Rev.}\ }\textbf {\bibinfo {volume}
  {168}},\ \bibinfo {pages} {1620} (\bibinfo {year} {1968})}\BibitemShut
  {NoStop}%
\bibitem [{\citenamefont {Lautrup}\ and\ \citenamefont
  {De~Rafael}(1968)}]{Lautrup:1968tdb}%
  \BibitemOpen
  \bibfield  {author} {\bibinfo {author} {\bibfnamefont {B.~E.}\ \bibnamefont
  {Lautrup}}\ and\ \bibinfo {author} {\bibfnamefont {E.}~\bibnamefont
  {De~Rafael}},\ }\href {https://doi.org/10.1103/PhysRev.174.1835} {\bibfield
  {journal} {\bibinfo  {journal} {Phys. Rev.}\ }\textbf {\bibinfo {volume}
  {174}},\ \bibinfo {pages} {1835} (\bibinfo {year} {1968})}\BibitemShut
  {NoStop}%
\bibitem [{\citenamefont {Eidelman}\ and\ \citenamefont
  {Jegerlehner}(1995)}]{Eidelman:1995ny}%
  \BibitemOpen
  \bibfield  {author} {\bibinfo {author} {\bibfnamefont {S.}~\bibnamefont
  {Eidelman}}\ and\ \bibinfo {author} {\bibfnamefont {F.}~\bibnamefont
  {Jegerlehner}},\ }\href {https://doi.org/10.1007/BF01553984} {\bibfield
  {journal} {\bibinfo  {journal} {Z. Phys. C}\ }\textbf {\bibinfo {volume}
  {67}},\ \bibinfo {pages} {585} (\bibinfo {year} {1995})},\ \Eprint
  {https://arxiv.org/abs/hep-ph/9502298} {arXiv:hep-ph/9502298} \BibitemShut
  {NoStop}%
\bibitem [{\citenamefont {Gounaris}\ and\ \citenamefont
  {Sakurai}(1968)}]{Gounaris:1968mw}%
  \BibitemOpen
  \bibfield  {author} {\bibinfo {author} {\bibfnamefont {G.~J.}\ \bibnamefont
  {Gounaris}}\ and\ \bibinfo {author} {\bibfnamefont {J.~J.}\ \bibnamefont
  {Sakurai}},\ }\href {https://doi.org/10.1103/PhysRevLett.21.244} {\bibfield
  {journal} {\bibinfo  {journal} {Phys. Rev. Lett.}\ }\textbf {\bibinfo
  {volume} {21}},\ \bibinfo {pages} {244} (\bibinfo {year} {1968})}\BibitemShut
  {NoStop}%
\bibitem [{\citenamefont {Kuhn}\ and\ \citenamefont
  {Santamaria}(1990)}]{Kuhn:1990ad}%
  \BibitemOpen
  \bibfield  {author} {\bibinfo {author} {\bibfnamefont {J.~H.}\ \bibnamefont
  {Kuhn}}\ and\ \bibinfo {author} {\bibfnamefont {A.}~\bibnamefont
  {Santamaria}},\ }\href {https://doi.org/10.1007/BF01572024} {\bibfield
  {journal} {\bibinfo  {journal} {Z. Phys. C}\ }\textbf {\bibinfo {volume}
  {48}},\ \bibinfo {pages} {445} (\bibinfo {year} {1990})}\BibitemShut
  {NoStop}%
\bibitem [{\citenamefont {Guerrero}\ and\ \citenamefont
  {Pich}(1997)}]{Guerrero:1997ku}%
  \BibitemOpen
  \bibfield  {author} {\bibinfo {author} {\bibfnamefont {F.}~\bibnamefont
  {Guerrero}}\ and\ \bibinfo {author} {\bibfnamefont {A.}~\bibnamefont
  {Pich}},\ }\href {https://doi.org/10.1016/S0370-2693(97)01070-8} {\bibfield
  {journal} {\bibinfo  {journal} {Phys. Lett. B}\ }\textbf {\bibinfo {volume}
  {412}},\ \bibinfo {pages} {382} (\bibinfo {year} {1997})},\ \Eprint
  {https://arxiv.org/abs/hep-ph/9707347} {arXiv:hep-ph/9707347} \BibitemShut
  {NoStop}%
\bibitem [{\citenamefont {Colangelo}\ \emph {et~al.}(2019)\citenamefont
  {Colangelo}, \citenamefont {Hoferichter},\ and\ \citenamefont
  {Stoffer}}]{Colangelo:2018mtw}%
  \BibitemOpen
  \bibfield  {author} {\bibinfo {author} {\bibfnamefont {G.}~\bibnamefont
  {Colangelo}}, \bibinfo {author} {\bibfnamefont {M.}~\bibnamefont
  {Hoferichter}},\ and\ \bibinfo {author} {\bibfnamefont {P.}~\bibnamefont
  {Stoffer}},\ }\href {https://doi.org/10.1007/JHEP02(2019)006} {\bibfield
  {journal} {\bibinfo  {journal} {JHEP}\ }\textbf {\bibinfo {volume} {02}},\
  \bibinfo {pages} {006}},\ \Eprint {https://arxiv.org/abs/1810.00007}
  {arXiv:1810.00007 [hep-ph]} \BibitemShut {NoStop}%
\bibitem [{\citenamefont {Melnikov}(2001)}]{Melnikov:2001uw}%
  \BibitemOpen
  \bibfield  {author} {\bibinfo {author} {\bibfnamefont {K.}~\bibnamefont
  {Melnikov}},\ }\href {https://doi.org/10.1142/S0217751X01005602} {\bibfield
  {journal} {\bibinfo  {journal} {Int. J. Mod. Phys. A}\ }\textbf {\bibinfo
  {volume} {16}},\ \bibinfo {pages} {4591} (\bibinfo {year} {2001})},\ \Eprint
  {https://arxiv.org/abs/hep-ph/0105267} {arXiv:hep-ph/0105267} \BibitemShut
  {NoStop}%
\bibitem [{\citenamefont {Jegerlehner}(2017)}]{Jegerlehner:2017gek}%
  \BibitemOpen
  \bibfield  {author} {\bibinfo {author} {\bibfnamefont {F.}~\bibnamefont
  {Jegerlehner}},\ }\href {https://doi.org/10.1007/978-3-319-63577-4} {\emph
  {\bibinfo {title} {{The Anomalous Magnetic Moment of the Muon}}}},\ Vol.\
  \bibinfo {volume} {274}\ (\bibinfo  {publisher} {Springer},\ \bibinfo
  {address} {Cham},\ \bibinfo {year} {2017})\BibitemShut {NoStop}%
\bibitem [{\citenamefont {Gonz\`alez-Sol\'\i{}s}\ and\ \citenamefont
  {Roig}(2019)}]{Gonzalez-Solis:2019iod}%
  \BibitemOpen
  \bibfield  {author} {\bibinfo {author} {\bibfnamefont {S.}~\bibnamefont
  {Gonz\`alez-Sol\'\i{}s}}\ and\ \bibinfo {author} {\bibfnamefont
  {P.}~\bibnamefont {Roig}},\ }\href
  {https://doi.org/10.1140/epjc/s10052-019-6943-9} {\bibfield  {journal}
  {\bibinfo  {journal} {Eur. Phys. J. C}\ }\textbf {\bibinfo {volume} {79}},\
  \bibinfo {pages} {436} (\bibinfo {year} {2019})},\ \Eprint
  {https://arxiv.org/abs/1902.02273} {arXiv:1902.02273 [hep-ph]} \BibitemShut
  {NoStop}%
\bibitem [{\citenamefont {Davier}\ \emph {et~al.}(2010)\citenamefont {Davier},
  \citenamefont {Hoecker}, \citenamefont {Lopez~Castro}, \citenamefont
  {Malaescu}, \citenamefont {Mo}, \citenamefont {Toledo~Sanchez}, \citenamefont
  {Wang}, \citenamefont {Yuan},\ and\ \citenamefont {Zhang}}]{Davier:2010fmf}%
  \BibitemOpen
  \bibfield  {author} {\bibinfo {author} {\bibfnamefont {M.}~\bibnamefont
  {Davier}}, \bibinfo {author} {\bibfnamefont {A.}~\bibnamefont {Hoecker}},
  \bibinfo {author} {\bibfnamefont {G.}~\bibnamefont {Lopez~Castro}}, \bibinfo
  {author} {\bibfnamefont {B.}~\bibnamefont {Malaescu}}, \bibinfo {author}
  {\bibfnamefont {X.~H.}\ \bibnamefont {Mo}}, \bibinfo {author} {\bibfnamefont
  {G.}~\bibnamefont {Toledo~Sanchez}}, \bibinfo {author} {\bibfnamefont
  {P.}~\bibnamefont {Wang}}, \bibinfo {author} {\bibfnamefont {C.~Z.}\
  \bibnamefont {Yuan}},\ and\ \bibinfo {author} {\bibfnamefont
  {Z.}~\bibnamefont {Zhang}},\ }\href
  {https://doi.org/10.1140/epjc/s10052-009-1219-4} {\bibfield  {journal}
  {\bibinfo  {journal} {Eur. Phys. J. C}\ }\textbf {\bibinfo {volume} {66}},\
  \bibinfo {pages} {127} (\bibinfo {year} {2010})},\ \Eprint
  {https://arxiv.org/abs/0906.5443} {arXiv:0906.5443 [hep-ph]} \BibitemShut
  {NoStop}%
\bibitem [{\citenamefont {Davier}\ \emph {et~al.}(2020)\citenamefont {Davier},
  \citenamefont {Hoecker}, \citenamefont {Malaescu},\ and\ \citenamefont
  {Zhang}}]{Davier:2019can}%
  \BibitemOpen
  \bibfield  {author} {\bibinfo {author} {\bibfnamefont {M.}~\bibnamefont
  {Davier}}, \bibinfo {author} {\bibfnamefont {A.}~\bibnamefont {Hoecker}},
  \bibinfo {author} {\bibfnamefont {B.}~\bibnamefont {Malaescu}},\ and\
  \bibinfo {author} {\bibfnamefont {Z.}~\bibnamefont {Zhang}},\ }\href
  {https://doi.org/10.1140/epjc/s10052-020-7792-2} {\bibfield  {journal}
  {\bibinfo  {journal} {Eur. Phys. J. C}\ }\textbf {\bibinfo {volume} {80}},\
  \bibinfo {pages} {241} (\bibinfo {year} {2020})},\ \bibinfo {note} {[Erratum:
  Eur.Phys.J.C 80, 410 (2020)]},\ \Eprint {https://arxiv.org/abs/1908.00921}
  {arXiv:1908.00921 [hep-ph]} \BibitemShut {NoStop}%
\end{thebibliography}%
\bibliographystyle{apsrev4-2}

\end{document}